\documentclass[11pt,a4paper]{article}
\usepackage{jheppub}
\usepackage{verbatim}
\usepackage{graphicx}
\usepackage{hyperref}
\usepackage{color}
\usepackage{amsmath}
\usepackage{tikz}
\usepackage{xcolor}
\usepackage{comment}
\usepackage{caption}
\usepackage{subcaption}
\usepackage{cleveref}
\usepackage[normalem]{ulem}

\usepackage[OT2,T1]{fontenc}
\DeclareSymbolFont{cyrletters}{OT2}{wncyr}{m}{n}

\title{Multicriticality in Yang-Lee edge singularity}
 
\author[a,b,c]{M\'at\'e Lencs\'es,}
\author[d,e]{Alessio Miscioscia,}
\author[f]{Giuseppe Mussardo,}
\author[a,b,g]{G\'abor Tak\'acs}

\affiliation[a]{Department of Theoretical Physics, Institute of Physics,
Budapest University of Technology and Economics, H-1111 Budapest, M{\H u}egyetem rkp. 3.}
\affiliation[b]{BME-MTA Momentum Statistical Field Theory Research Group, Institute of Physics,
Budapest University of Technology and Economics, H-1111 Budapest, M{\H u}egyetem rkp. 3.}
\affiliation[c]{Wigner Research Centre for Physics,
Konkoly-Thege Miklós u. 29-33, 1121 Budapest , Hungary}
\affiliation[d]{Dipartimento di Fisica e Astronomia, Universit\' a degli Studi di Padova, via Marzolo 8, 35131 Padova, Italy}
\affiliation[e]{Deutsches Elektronen-Synchrotron DESY, Notkestr. 85, 22607 Hamburg, Germany}
\affiliation[f]{SISSA \& INFN, Sezione di Trieste, via Bonomea 265, I-34136, Trieste, Italy}
\affiliation[g]{MTA-BME Quantum Correlations Group (ELKH), Institute of Physics,
Budapest University of Technology and Economics, H-1111 Budapest, M{\H u}egyetem rkp. 3.}

\emailAdd{alessio.miscioscia@desy.de}

\newcommand{\op}[1]{\boldsymbol{#1}}
\newcommand{\de}{\mathrm d}
\newcommand{\Tr}{\operatorname{Tr}}
\newcommand{\ket}[1]{\left |#1 \right \rangle }

\begin{document}

\tikzset{every picture/.style={line width=0.75pt}} 
\begin{flushright}
DESY-22-162 \\
\end{flushright}
\abstract{
In this paper we study the non-unitary deformations of the two-dimensional Tricritical Ising Model obtained by coupling its two spin $\mathbb{Z}_2$ odd operators to imaginary magnetic fields. Varying the strengths of these imaginary magnetic fields and adjusting correspondingly the coupling constants of the two spin $\mathbb{Z}_2$ even fields, we establish the presence of two universality classes of infrared fixed points on the critical surface. The first class corresponds to the familiar Yang-Lee edge singularity, while the second class to its tricritical version. We argue that these two universality classes are controlled by the conformal non-unitary minimal models $\mathcal M(2,5)$ and $\mathcal M(2,7)$ respectively, which is supported by considerations based on $\op P \op T$ symmetry and the corresponding extension of Zamolodchikov's $c$-theorem, and also verified numerically using the truncated conformal space approach. Our results are in agreement with a previous numerical study of the lattice version of the Tricritical Ising Model \cite{vonGehlen:1994rp}. We also conjecture the classes of universality corresponding to higher non-unitary multicritical points obtained by perturbing the conformal unitary models with imaginary coupling magnetic fields.
}


\date{17th October 2022}
\maketitle
\flushbottom

\section{Introduction}
In the modern study of quantum field theories a crucial role is played by the Renormalisation Group (RG) which provides a systematic approach to mapping the space of quantum field theories. Indeed, starting from an RG fixed point and deforming it with some relevant fields, one can follow the RG flow from the original ultraviolet fixed point to either a massive quantum field theory or a nontrivial infrared fixed point. 

Under some quite general assumptions, the fixed points of the RG flows in two space-time dimensions are described by conformal field theories (CFTs) with a symmetry algebra that contains the infinite dimensional Virasoro algebra. The simplest CFTs are the \textit{minimal models} $\mathcal M(p,q)$ indexed by two positive co-prime integers $2\leq p<q$, with a Hilbert space built from a finite number of irreducible representations of the Virasoro algebra \cite{Belavin:1984vu}. The central charge of these models is given by 
\begin{equation}
c \,=\, 1 - \frac{6 (p-q)^2}{p q }\,\,\,.
\label{centralcharge}
\end{equation}
Models with $q=p+1$ are unitary, and for the case of  diagonal modular invariant partition functions \cite{Cardy:1986ie,Cappelli:1986hf} they are in a one-to-one correspondence with multicritical Ising models described by the $\varphi^{2(p-1)}$ Landau--Ginzburg Lagrangians \cite{Zamolodchikov:1986db}. The CFT $\mathcal M(3,4)$ describes the \textit{Critical Ising} universality class, $\mathcal M(4,5)$ corresponds to \textit{Tricritical Ising}, and in general the model $\mathcal M(p+1,p+2)$ describes the $p$th order multicritical Ising fixed point.

What is the physical interpretation of the non-unitary minimal models in the framework of statistical mechanics of critical phenomena? The answer to this question is not known in general. Nevertheless, the simplest of these models $\mathcal M(2,5)$  describes the class of universality of the Yang--Lee edge singularity of the zeros of the grand canonical partition function of the Ising model \cite{Yang:1952be,Lee:1952ig, Fisher:1978pf,Cardy:1985yy}. The Yang-Lee minimal model $\mathcal M(2,5)$ has been recently studied from a RG flow perspective, i.e. as the endpoint of a non-unitary RG flow which connects the Ising model in the ultraviolet and the Yang-Lee model in the infrared \cite{Fonseca:2001dc,Xu:2022mmw}, where the flow is induced by coupling the Ising model to an imaginary magnetic field and tuning the coupling of the energy density operator of the model appropriately. 
Adopting the same RG approach, in this paper we argue that the series of non-unitary minimal models  $\mathcal M(2, 2 n +3)$ $(n=1, 2, \ldots)$ control the multi-criticality of Yang-Lee zeros associated to imaginary magnetic perturbations of the unitary minimal models $\mathcal M(p,p+1)$. In particular, we discuss in great detail the emergence of tricriticality in the Yang-Lee zeros starting from the $\mathcal M(4,5)$  describing the class of universality of the Tricritical Ising Model, associated to the $\varphi^6$ Landau-Ginzburg Lagrangian \cite{Friedan:1983xq,Friedan:1984rv}. 

The layout of the paper is as follows. In Section \ref{sec:LY_zeros} we briefly recall the theory of the Yang-Lee zeros and the edge singularity, while in Section \ref{sec:TCSA_approach} we describe briefly the application of the Truncated Conformal Space Approach (TCSA) to the study of non-unitary fixed points. As a warm-up we consider the standard Yang-Lee criticality in the TCSA in Section \ref{sec:IFT_YL}, and then turn to the case of the Tricritical Ising Model in Section \ref{sec:TIM_LY}, concluding in Section \ref{sec:Conclusions}. The paper also has an Appendix which presents some considerations regarding the Landau-Ginzburg formulation of the non-unitary minimal models $\mathcal M(2, 2 n +3)$.

\section{Yang-Lee zeros and edge singularity}\label{sec:LY_zeros}

The study of the Yang-Lee zeros started with the seminal works by Yang and Lee \cite{Yang:1952be,Lee:1952ig} who realised that the analytic properties of the density of a gas in the thermodynamic limit is determined by the behaviour of the zeros of the grand canonical partition function considered in the complex plane of the fugacity. Let's consider the grand canonical partition function of the finite volume model where $M$ is the maximum number of particles that can be contained in the volume $V$ 
\begin{equation}
\Omega_V(z) = \sum_{N = 0}^M \frac{\mathcal Z_N(V)}{N!}z^N  = \prod_{l = 1}^M\left(1-\frac{z}{z_l}\right )\ ,
\end{equation}
with $z$ denoting fugacity in terms of which $\Omega_V$ has (complex) zeros $z_1,\ldots, z_M$, usually called \textit{Yang-Lee zeros}. In the thermodynamic limit $V \to \infty$, the pressure and the density can be expressed as
\begin{align}
	\frac{p(z)}{k T} &= \lim_{V \to \infty}\frac{\ln \Omega_V(z)}{V} = \lim_{V \to \infty}\frac{1}{V}\sum_{l = 1}^M  \ln \left(1-\frac{z}{z_l}\right )\,, \nonumber\\ 
\rho (z) &= \lim_{V \to \infty}z \frac{\de}{\de z}\frac{\ln \Omega_V(z)}{V} = \lim_{V \to \infty }\frac{1}{V}\sum_{l = 1}^M \frac{z}{z-z_l}  \ .\label{eq:TDlimit}
\end{align}
For extended systems the limit $V \to \infty$ also enforces $M \to \infty$ and therefore the number of zeros becomes infinite, with their location in the complex plane described by a density function $\eta(z)$ related to the pressure and other physical quantities by an integral form of the relations \eqref{eq:TDlimit} \cite{Mussardo:2017gao}. Phase transitions occur when there are zeros which pinch the positive real axis\cite{Yang:1952be,Lee:1952ig}. 

Using the well-known mapping between density and magnetisation in such binary systems (see for instance \cite{Mussardo:2020rxh}), the same holds for Ising magnets when the magnetisation is extended to the complex plane.  The partition function of a system on a lattice of $L\times N$ sites can be written as the trace of $L$-th power of the \textit{transfer matrix} ${\mathcal T}$
\begin{equation}
    \Omega_L = \Tr \left(e^{-\beta \op H}\right) = \Tr \left(\op{\mathcal T}^L\right) = \lambda_1^L+\lambda_2^L+\ldots \lambda_N^L \ ,
\end{equation} 
where $\lambda_1,\dots,\lambda_N$ are eigenvalues of ${\mathcal T}$.

The condition for a phase transition to occur can be phrased in terms of the behaviour of the eigenvalues $\lambda_1,\dots,\lambda_N$, which depend on the couplings $g_1,\ldots, g_n$ of the systems entering the Hamiltonian $\op H$. The free energy per site in the thermodynamic limit reads 
\begin{equation}
    f(g_1,\ldots, g_n) = \ln\left( \max\{\lambda_i(g_1,\ldots, g_n)\}\right) \,,
\end{equation}
and it is evident that non-analyticities in the free energy may occur at crossing of the eigenvalues, most significantly of the largest two of them, which has been explicitly and analytically verified for the Ising model \cite{Deger:2020mdv}. This means that a phase transition is expected when the couplings $g_1,\ldots, g_n$ are set to produce an eigenvalue crossing. It is important to notice that in general, when there are many relevant fields, it is not sufficient to tune only the temperature and the magnetic field to implement this condition. 

The case most extensively discussed in literature is that of the Ising model, with a lattice Hamiltonian given by 
\begin{equation}
	 \op H = J \sum_{\langle i,j \rangle}s_i s_j + h \sum_{i}s_i \ , \hspace{1 cm}s_i \in \{-1,1\} \ ,
\end{equation} 
where $\langle i,j\rangle$ indicates the sum over nearest neighbour sites. 
For this particular model it was established that the zeros of the partition function lie on a unit circle in the complex plane of the variable $z = e^{-\beta h} = e^{i \theta}$ (here $\beta = (k T)^{-1}$ is the inverse temperature and $h$ is the external magnetic field) \cite{Lee:1952ig}. When the temperature $T$ is larger then the critical value $T_c$, the zeros are located on a symmetric arc centred around $\theta=i \beta h = \pi$ with edges are at $\pm \theta_0$. These two edges pinch the real axis for $T = T_c$ and the circle remains closed for  $T<T_c$ (see Fig. 1). 

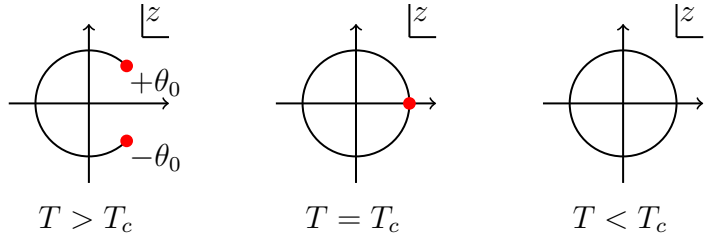
\begin{figure}[h]

\centering
\begin{tikzpicture}[x=0.5pt,y=0.5pt,yscale=-1,xscale=1]
\draw[black, ->] (-60,0) -- (60,0);
\draw[black, ->] (0,60) -- (0,-60);
\draw[black] (60,-50)--(40,-50);
\draw[black] (40,-75)--(40,-50);
\draw (40,-75) node [anchor=north west][inner sep=0.75pt]  [font=\large]  {$z$};

\draw[black, ->] (-260,0) -- (-140,0);
\draw[black, ->] (-200,60) -- (-200,-60);
\draw[black] (-140,-50)--(-160,-50);
\draw[black] (-160,-75)--(-160,-50);
\draw (-160,-75) node [anchor=north west][inner sep=0.75pt]  [font=\large]  {$z$};

\draw[black, ->] (140,0) -- (260,0);
\draw[black, ->] (200,60) -- (200,-60);
\draw[black] (260,-50)--(240,-50);
\draw[black] (240,-75)--(240,-50);
\draw (240,-75) node [anchor=north west][inner sep=0.75pt]  [font=\large]  {$z$};

\draw[color=black](0,0) circle (40);
\filldraw[red] (40,0) circle (2pt) node[anchor=north]{};

 \draw[black] (-171.72,28.28) arc (45:315:40);
\filldraw[red] (-171.72,28.28) circle (2pt) node[anchor=north]{};
\filldraw[red] (-171.72,-28.28) circle (2pt) node[anchor=north]{};

\draw[color=black](200,0) circle (40);

\draw (-240,75) node [anchor=north west][inner sep=0.75pt]  [font=\large]  {$T> T_c$};

\draw (160,75) node [anchor=north west][inner sep=0.75pt]  [font=\large]  {$T< T_c$};

\draw (-40,75) node [anchor=north west][inner sep=0.75pt]  [font=\large]  {$T= T_c$};

\draw (-171.72,28.28) node [anchor=north west][inner sep=0.75pt]  [font=\large]  {$-\theta_0$};
\draw (-171.72,-28.28) node [anchor=north west][inner sep=0.75pt]  [font=\large]  {$+\theta_0$};

\end{tikzpicture}
\caption{Distribution of the Yang-Lee zeros for the Ising model in the complex plane of the fugacity $z$. } \label{Fig.IsingZeros}
\end{figure}
Kortman and Griffiths pointed out that the density of the zeros $\eta(\theta)$ has an anomalous behaviour nearby the two edges \cite{Kortman:1971zz} given by 
\begin{equation} 
\eta (\theta) \overset{\theta\to \theta_0}{\sim} |\theta-\theta_0|^\mu \ , 
\end{equation}
with a negative value for $\mu$. Since $e^{-\beta h} = e^{i \theta}$, the density of the zeros can be written as a function of the magnetic field and there must be an imaginary value of the magnetic field $ih_0$ for which the density is divergent. As a result the magnetisation is also divergent for $h=i h_0$, which is equivalent to a phase transition. Fisher \cite{Fisher:1978pf} proposed to study this critical phenomenon from a field theoretical point of view according to the following lines\footnote{Here we recall Fisher's argument as reformulated by von Gehlen \cite{vonGehlen:1994rp}.}. Starting from the Landau-Ginzburg Lagrangian of the Ising model is 
\begin{equation}
\mathcal L_{Ising \, L.G.} = \frac{1}{2} \partial_\mu \varphi \partial^\mu \varphi + g_1 \varphi+ g_2 \varphi^2+ \varphi^4 \ , 
\end{equation}
(with the critical point defined by $g_1 = g_2  = 0$), the field theoretic formulation of the Yang-Lee edge singularity can be obtained by shifting the order parameter with an imaginary constant term and then tuning $g_2$ to set the quadratic term to zero to preserve criticality \cite{Fisher:1978pf}. Denoting the magnetic field $g_1$ by $h$, this results in the Lagrangian 
\begin{equation}\label{YLLagrangian}
\mathcal L_{YL} = \frac{1}{2} \partial_\mu \varphi \partial^\mu \varphi + (h-i h_c)  \varphi + i \gamma \varphi^3 + \ldots \ .
\end{equation}
The Lagrangian \eqref{YLLagrangian} is the effective Landau-Ginzburg description of the Yang-Lee critical point: exactly at the critical value of the imaginary magnetic field $h= i h_c$ it is a cubic Lagrangian with an imaginary coupling. Using Fisher's result, it was shown later by Cardy that the conformal field theory corresponding to this class of universality in two dimensions is the minimal model $\mathcal M(2,5)$ \cite{Cardy:1985yy}. His argument was based on the fact that the Lagrangian \eqref{YLLagrangian} allows only one relevant field (because $\varphi^2$ is related to $\varphi$ via equations of motion, cf. Eq. (\ref{phii3}) below), that the three point function of the resulting theory cannot be zero and on the assumption (that can be checked a posteriori) that the resulting theory is a (non-unitary) minimal model. We do not review Cardy's argument in detail here; we limit ourselves to observe that Cardy's finding is consistent with the Landau-Ginzburg point of view, which can be adapted to some non-unitary cases. The simplest case is just the minimal model $\mathcal M(2,5)$ that contains only two fields: the identity $\op 1 $ and a field $\phi$ of weights $(\Delta_\phi, \overline \Delta_\phi) = (-1/5,-1/5)$. For this reason the minimal model $\mathcal M(2,5)$ is very well studied from both analytic and numerical point of view \cite{Cardy:1985yy,Itzykson:1986pk,Yurov:1989yu,Cardy:1989fw,Zamolodchikov:1990bk}. The only nontrivial operator product expansion (OPE) is given by
\begin{equation}
\phi(x) \phi (x') = |x-x'|^{4/5} (\op 1 + \text{descendants})+c_{\phi\phi}^\phi |x-x'|^{2/5} (\phi(x) + \text{descendants}) \ ,
\label{eq:YL_OPE}\end{equation}
where the structure constant is
\begin{equation}
 c_{\phi\phi}^\phi = i \left(\frac{\Gamma(1/5)}{\Gamma(4/5)}\right)^{3/2}\left(\frac{\Gamma(2/5)}{\Gamma(3/5)}\right)^{1/2}  \ .
\end{equation}
From the OPE \eqref{eq:YL_OPE} we are led directly to the equation of motion 
\begin{equation}
	 \op L_{-1}\overline {\op L}_{-1}\phi=\partial\bar{\partial}\phi \sim i : \phi^2 : \ ,
\label{phii3}
\end{equation}
which corresponds to the Landau-Ginzburg \eqref{YLLagrangian} at the critical point, 
where $\op L_{-1}$ and $\overline {\op L}_{-1}$ are Virasoro generators of chiral translations, cf. Eqs. (\ref{Virasoro},\ref{overVirasoro}) below. 

To sum up, based on the results by Fisher and Cardy we know that the CFT behind the critical behaviour of the Yang-Lee zeros in the Ising case is the minimal model $\mathcal M(2,5)$.  Furthermore the field theoretical point of view suggests how we can reach the minimal model $\mathcal M(2,5)$ from an ultraviolet unitary theory. Indeed starting from the critical Ising fixed point perturbed with a purely imaginary magnetic field and suitably adjusting the coupling in front of the energy operator of the model, we expect to find a critical point corresponding to the minimal model $\mathcal M(2,5)$. In the RG language this means that starting from the critical Ising as UV fixed point and following a suitable RG flow we expect to reproduce the minimal model $\mathcal M(2,5)$ in the infrared. Let us comment that, even if in the Ising case only two couplings (the temperature and the magnetic field) appear, for more complicated models other relevant couplings must also be taken into account. Indeed, in general the physical magnetic field is a combination of all the odd fields and the full scaling region, i.e. all the relevant deformations, has to be taken in consideration. To make further progress, let's first briefly recall the numerical method we use to follow the RG flows.

\section{Truncated conformal space approach and finite size spectrum}\label{sec:TCSA_approach}

\subsection{The truncated conformal space approach}
The main tool we are going to use is the \textit{Truncated Conformal Space Approach} (TCSA), originally proposed by Yurov and Zamolodchikov \cite{Yurov:1989yu}. The idea behind this approach is rather simple and based on the fact that on a separable Hilbert space a generic Hamiltonian  $\op H$ can be expressed in terms of the matrix elements computed in a suitably ordered basis and then truncated to the first $N$ energy levels. Increasing the number $N$ of eigenstates this approach gives access to the non-perturbative spectrum of the lowest lying levels to a certain approximation.

Since in our analysis we start from the ultraviolet fixed points of the RG flow (chosen to be a conformal minimal model), the natural choice for the computational basis is that of the CFT eigenstates of the ultraviolet fixed point. The two-dimensional CFT is invariant under the left/right Virasoro algebras
\begin{equation}
\label{Virasoro}
[\op L_n,\op L_m] = (n-m) \op L_{n+m}+\frac{c}{12} n \left (n^2-1\right )\delta_{n+m,0} \ ,
\end{equation}
\begin{equation}
\label{overVirasoro}
[\overline {\op L}_n,\overline {\op L}_m] = (n-m) \overline{\op L}_{n+m}+\frac{c}{12} n \left (n^2-1\right )\delta_{n+m,0} \ ,
\end{equation}
\begin{equation}
[\op L_n,\overline {\op L}_m] = 0\ ,
\end{equation}
where $c$ is the \textit{central charge} characterizing the CFT. Invariance under the Virasoro algebra allows us to factorise the Hilbert space, $\mathcal  H$, in terms of tensor products of irreducible representations $V(\phi)$ and $\overline V(\phi)$ of the left/right Virasoro algebras\footnote{Here we consider only conformal field theories corresponding to the diagonal modular invariant torus partition function \cite{Cappelli:1986hf}, since for the minimal models these are the ones that describe the multicritical Ising universality classes.} 
\begin{equation}
    \mathcal H =  \bigoplus_{\phi} V(\phi)\otimes \overline V(\phi) \ ,
\label{eq:CFT_Hilbert}
\end{equation} 
where the fields $\phi$ corresponds to a primary state $\ket{\phi}$, i.e. a state that is annihilated by $\op L_n$ where $n>0$. The irreducible representation can be constructed starting from the Verma module spanned by the states
\begin{equation}
\ket{\phi; n_1, n_2,\ldots, n_k} = \op L_{-n_1}\op L_{-n_2}\ldots \op L_{-n_k} \ket{\phi} \ , \hspace{1 cm} n_1 \le n_2 \le \ldots \le n_k \ ,
\end{equation}
which is, however, generally reducible and it is necessary to eliminate singular vectors and their descendants. The sum $n_1+n_2+\ldots+n_k$ is called the (left) level of the descendant state, with right descendants and their levels defined similarly using the $\overline {\op L}$ operators.

For minimal conformal models with $c<1$ the number of distinct irreducible representations that appear in the space of states is finite\cite{Belavin:1984vu} and the Hilbert space can be obtained as a combination of a finite number of terms in \eqref{eq:CFT_Hilbert}. The possibilities are restricted by modular invariance and have been classified in \cite{Cappelli:1986hf}. 

\begin{figure}
\centering
\begin{tikzpicture}[x=0.5pt,y=0.5pt,yscale=-1,xscale=1]

\draw   (256.56,1308.44) .. controls (256.56,1297.4) and (289.56,1288.44) .. (330.28,1288.44) .. controls (370.99,1288.44) and (404,1297.4) .. (404,1308.44) .. controls (404,1319.49) and (370.99,1328.44) .. (330.28,1328.44) .. controls (289.56,1328.44) and (256.56,1319.49) .. (256.56,1308.44) -- cycle ;
\draw    (256.56,1308.44) -- (257.56,1433.67) ;
\draw [shift={(257.06,1371.06)}, rotate = 89.54] [fill={rgb, 255:red, 0; green, 0; blue, 0 }  ][line width=0.08]  [draw opacity=0] (10.72,-5.15) -- (0,0) -- (10.72,5.15) -- (7.12,0) -- cycle    ;
\draw    (404,1308.44) -- (404.56,1432.89) ;
\draw    (257.56,1433.67) .. controls (268.11,1458.89) and (400.11,1458.89) .. (404.56,1432.89) ;
\draw [shift={(332.3,1452.49)}, rotate = 180.68] [fill={rgb, 255:red, 0; green, 0; blue, 0 }  ][line width=0.08]  [draw opacity=0] (10.72,-5.15) -- (0,0) -- (10.72,5.15) -- (7.12,0) -- cycle    ;
\draw  [dash pattern={on 0.84pt off 2.51pt}]  (257.56,1433.67) .. controls (267.11,1406.89) and (391.11,1406.89) .. (404.56,1432.89) ;

\draw (334,1308.89)--(404,1308.89);

\draw (364,1265.44) node [anchor=north west][inner sep=0.75pt]  [font=\large]  {$R$};
\draw (334,1460.44) node [anchor=north west][inner sep=0.75pt]  [font=\large]  {$x$};
\draw (234,1365.44) node [anchor=north west][inner sep=0.75pt]  [font=\large]  {$y$};
\end{tikzpicture}
\caption{Space-time cylinder of circumference $R$.}
\label{cylinder}
\end{figure}
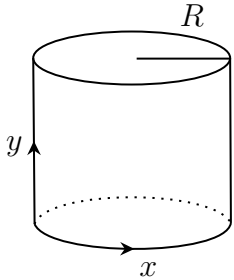

For a theory defined on a cylinder of circumference $R$ as shown in Fig. \ref{cylinder}, the CFT Hamiltonian $\op H_{CFT}$ has the form  \begin{equation}
    \op H_{CFT} = \frac{2 \pi}{R} \left(\op L_0 + \op {\overline L}_0-\frac{c}{12}\right) \ ,
\end{equation}
which is diagonal in the conformal basis. Adding to the relevant deformation, the full Hamiltonian reads \begin{equation}
	\op H = \op H_{CFT}+ \op V = \op H_{CFT}+ \lambda \int_0^R \phi \ \de x \ ,
\end{equation}
where $\lambda$ is a coupling constant and $\phi$ is one of the relevant fields of the ultraviolet CFT, with conformal weights $\Delta=\overline \Delta$. Here for simplicity we only included a single perturbing field, but the approach can be easily extended to an arbitrary number of perturbations. In the conformal basis the matrix elements of the perturbing field can be computed iteratively using the structure constants of the primary fields and the explicit expression of the descendant fields of the Verma modules. Several computer programs have been developed to this aim starting with \cite{Lassig:1990cq}; here we used a recent advanced implementation of the TCSA which exploits the chiral factorisation of the Hilbert space for increased efficiency~\cite{Horvath:2022zwx}. 

Due to the presence of the perturbing term, the Hamiltonian is no longer diagonal, but schematically the dependence of the Hamiltonian on the circumference of the cylinder is given by \begin{equation}
 \op  H = \frac{2 \pi}{R} \begin{pmatrix}
\star & 0 & 0 &\dots  &\dots   \\ 
0 &\star  &0  &\dots & \dots   \\ 
0 &0  & \star  & \ddots & \ddots   \\ 
\vdots & \vdots & \ddots &\ddots & \ddots   \\
\end{pmatrix}+ R^{1-2\Delta}\begin{pmatrix}
\star & \star & \star &\dots  &\dots   \\ 
\star &\star  &\star  &\dots & \dots   \\ 
\star &\star  & \star  & \ddots & \ddots   \\ 
\vdots & \vdots & \ddots &\ddots & \ddots   \\
\end{pmatrix} \ .
\end{equation}
The Hilbert space can now be truncated to states that correspond to conformal energy less then a chosen cut-off value $\Lambda$. Given that the eigenvalues of $\op L_0$ read 
\begin{equation}
\op L_0 \ket{\phi; n_1,\ldots, n_k}  = \Delta_\phi+ n_1+\ldots + n_k \ ,
\end{equation}
and similarly for $\overline {\op L}_0$, one keeps the states that satisfy 
\begin{equation}
	\frac{2 \pi}{R}(\Delta_\phi+\overline \Delta_\phi+n_1+\overline n_1+\ldots+ n_k+\overline n_k) \le \Lambda \ .
\end{equation}
Instead of the physical cut-off parameter $\Lambda$ it is more convenient to truncate the Hilbert space by introducing a level cut-off, i.e. an upper limit on the descendant level of the states retained:
\begin{equation}
	n_1+\ldots+ n_k\le N_\text{max} \ .
\end{equation}
Note that we specify the truncation in terms of the right chiral level since this automatically imposes an upper limit on the left chiral level in all momentum sectors (in particular, in the zero-momentum sectors the two descendant levels are eventually equal).

The truncated Hamiltonian is a finite square matrix which can be diagonalised numerically to find the non-perturbative spectrum up to a certain energy scale. For more details on TCSA specifically in the scaling tricritical Ising field theory we refer the interested reader to \cite{2022ScPP...12..162C}. Finally we note that the truncation dependence can be formulated in terms of a the renormalisation group with $\Lambda$ as the ultraviolet cutoff \cite{2008JSMTE..03..011F,2011arXiv1106.2448G,2015PhRvD..91h5011R,2015JHEP...09..146L}.

\subsection{Finite size spectrum near the critical points}

Let $\xi$ be the correlation length of the perturbed theory on the cylinder.  For $R \ll \xi$ the energy levels in zero momentum sector are expected to scale as
\begin{equation}
E_i\simeq \frac{2 \pi}{R}\left(2 \Delta^\text{UV} + 2 n^\text{UV}-\frac{c^\text{UV}}{12} \right) \ ,
\end{equation}
where $\Delta^\text{UV}$ and $n^\text{UV}$ are the primary conformal weight and the descendent level of the conformal eigenstate to which the energy level $E_i$ corresponds, while $c^\text{UV}$ is the central charge of the ultraviolet theory. In the opposite limit of large $R$ the energy spectrum is that of the infrared theory (including a bulk energy term), which depends on the nature of the infrared fixed point: RG flows to a massive QFT have a behaviour different from those ending in an IR fixed point described another conformal field theory. 

\begin{figure}[t]
\centering
\begin{tikzpicture}[x=0.6pt,y=0.6pt,yscale=-1,xscale=1]
\draw[dashed] (0,45)-- (50,45);
\draw[->] (50,45) -- (400,45);
\filldraw[red] (370,45) circle (4pt) node[anchor=north]{};
\draw (50,55) node [anchor=north west][inner sep=0.75pt]  [font=\Large]  {$R$};
\draw[dashed] (200,25) -- (340,25) -- (340,65) -- (200,65) -- (200,25);
\draw (200,0) node [anchor=north west][inner sep=0.75pt]  [font=\large]  {physical window};
\draw (345,55) node [anchor=north west][inner sep=0.75pt]  [font=\large]  {Critical point};
\end{tikzpicture}
\caption{The physical window is identified as a range of radius small enough that the truncation errors are negligible, but large enough that the infrared theory dominates the spectrum. The critical point (in red) is  located at $R \to \infty$.}
\label{PhysicalWindow}
\end{figure}
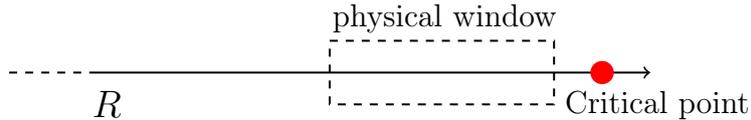

For massive flows, both the mass spectrum of the theory and the scattering phases can be extrapolated from the spectrum \cite{Luscher:1985dn,Luscher:1986pf} and many interesting examples of these computations are provided in the literature (e.g. \cite{Lassig:1990xy,Lassig:1990wc,Delfino:1996xp,Gabai:2019ryw}). However, for the purpose of our work we are mostly interested in massless flows which are characterised by the presence of a new critical point in the infrared. However, since the TCSA is always implemented in a finite volume, it is important to emphasise that one could not expect to reach this new critical point exactly due to the divergence of the correlation length. Nevertheless, it is possible to approach the critical point to extract its characteristics, although it is necessary to keep in mind that the larger is the volume, the greater the spectrum is afflicted by errors due to truncation. For this reason the right strategy is to identify a \textit{physical window}, as illustrated in Fig. \ref{PhysicalWindow}, i.e. a range of volumes and couplings, in which we can reasonably interpret the TCSA results in terms of the infrared massless theory. To do so, it is first necessary to attempt to identify the infrared critical point by locating the values of the couplings where the mass gap vanishes. Ideally this point would be reached for $R \to \infty$ but, since TCSA does not allow to study the spectrum at infinite volume, the asymptotic behaviour can be extracted by a proper choice of the physical window by looking for a volume range in which the relative energy levels scale as $1/R$. As a result, the critical values of the couplings obtained by TCSA generally depend on the choice of the cutoff $\Lambda$.

Once the physical window is identified, one expects that the large volume behaviour  reproduces the spectrum of the infrared conformal field theory plus a bulk energy 
term\footnote{Since we are not exactly at the critical point, one can study the infrared theory in terms of an the effective field theory around the infrared conformal field theory. This is the approach followed in the study of the Yang-Lee edge singularity \cite{Xu:2022mmw}.} ${\mathcal F}$
\begin{equation}
E_i\simeq \frac{2 \pi}{R}\left(2 \Delta^\text{IR} + 2 n^\text{IR}-\frac{c^\text{IR}}{12} \right)+ {\mathcal F} R \ .
\end{equation}
The bulk energy density ${\mathcal F}$ can be conveniently removed from the discussion by considering differences between energy levels. In particular, subtracting the ground state from each energy level gives
\begin{equation}\label{eq: Ci definition}
E_i-E_0 \simeq  \frac{4 \pi}{R} \left (\Delta^\text{IR}-\Delta_{min}^\text{IR}+n^\text{IR}\right ) = \frac{4 \pi}{R} C_{i} \ ,
\end{equation}
where in the physical window the $C_i$ are expected to be constant, with their values dictated by the infrared fixed point.

\section{Ising field theory and Yang-Lee edge }\label{sec:IFT_YL}

\begin{table}[t]
\begin{center}
\begin{tabular}{ c c c c}
\hline 
Primary & Weights & LG field & name \\ 
$\op 1$ & (0,0) &  $1$ & Identity \\
$\sigma$ & (1/16,1/16) & $\varphi$ & magnetisation \\
$\epsilon$ & (1/2,1/2) & $:\varphi^2:$ & energy \\
 \hline 
\end{tabular}
\caption{Primary fields in the critical Ising model $\mathcal M(3,4)$.}
\label{IMContent}
\end{center}
\end{table}

Here we consider the Ising field theory realised as a perturbation of the conformal minimal model $\mathcal M(3,4)$, namely the critical Ising CFT. The primaries of the Ising CFT are listed in Table \ref{IMContent}, and consists of two fields (besides the identity): the magnetic field $\sigma$ that could be identified as the order parameter of the corresponding Landau-Ginzburg description, and the energy field $\epsilon$ that is the renormalised square of the order parameter according to the correspondence between unitary minimal models and Landau-Ginzburg multicritical theories \cite{Zamolodchikov:1986db}.

Using the OPEs of the minimal model $\mathcal M(3,4)$ the Landau-Ginzburg Lagrangian of the Ising model takes the form (see appendix \ref{sec:AppendixA})\begin{equation}
\mathcal L_\text{Ising LG} = \frac{1}{2} \partial_\mu \varphi \partial^\mu \varphi + g_1 \varphi+ g_2 \varphi^2+ \varphi^4 \ .
\end{equation}
where the critical point is defined by $g_1 = g_2  = 0$.

\subsection{$\op P \op T$ symmetry and its breaking}
The CFT Hamiltonian of the Ising model $\mathcal M(3,4)$ can be written in terms of two fermionic fields $\psi$ and $\overline \psi$ of weights $(1/2,0)$ and $(0,1/2)$ respectively, as a theory of free Majorana fermions. Switching on a fermion mass corresponds to the displacement of the temperature $T$ from the critical value $T_c$, given as a deformation by the energy field which can be written in terms of the fermion fields as $\epsilon= 2 \pi i \overline \psi \psi$, while the magnetic field $\sigma$ is non-local with respect to the fermions. The (Euclidean) Lagrangian density of the scaling Ising field theory is given by
\begin{equation}\label{IsingFieldTheoryFermions}
\mathcal L_\text{IFT} = \psi \overline \partial \psi + \overline \psi \partial\overline  \psi + i m \overline \psi \psi + h \sigma \ ,
\end{equation}
where $\partial = \frac{1}{2}(\partial_x-i \partial_y)$ and $\overline \partial = \frac{1}{2}(\partial_x+i \partial_y)$  ($x$ and $y$ are the two coordinates of the space-time cylinder). The scaling region described by the Lagrangian \eqref{IsingFieldTheoryFermions} (with real couplings) was studied in details \cite{Fonseca:2006au,Zamolodchikov:2011wd,Zamolodchikov:2013ama}. 

To obtain the Yang-Lee singularity we consider the RG flow obtained when replacing $h \to i h$ resulting in the Lagrangian density 
\begin{equation}\label{IsingFermionLagr}
\mathcal{L}_\text{iIFT} = \psi \overline \partial \psi + \overline \psi \partial\overline  \psi + i m \overline \psi \psi + i h \sigma \ .
\end{equation}
The Hamiltonian associated to this Lagrangian is not Hermitian so that reality of the energy levels depends on the $\op P \op T$ symmetry of the system. Considering an eigenvector $\ket{\Psi}$ of the a non-Hermitian Hamiltonian ${\op H}$, three possible scenarios can be realised depending on the validity of the following two assumptions\footnote{Note that $\op P\op T$ is an anti-linear operator.} 
\begin{equation}
 i) \ \  [\op H, \op P \op T] =  0 \ , \hspace{1 cm} ii) \ \ \op P \op T \ket{\Psi} = e^{i \alpha} \ket{\Psi }\ , \ \ \ \alpha \in \mathbb R\ . 
\end{equation}
If neither $i)$ nor $ii)$ are satisfied, then the system is not $\op P \op T$ symmetric; if both $i)$ and $ii)$ hold, the system lies in a $\op P \op T$-symmetric phase;  if instead $i)$ holds but $ii)$ no longer holds, the system is in a spontaneously broken $\op P \op T$ phase. In the non-$\op P \op T$ symmetric regime the spectrum is in general complex, while in the $\op P \op T$-symmetric phase the spectrum is real \cite{Fring:2022tll}. In the spontaneously broken $\op P \op T$ regime,  $\ket{\Psi}$ and $\op P \op T \ket{\Psi}$ are two eigenvectors of the Hamiltonian with complex conjugate eigenvalues therefore all energies are either real or appear in complex conjugate pairs \cite{Fring:2022tll}.

\begin{figure}[t]
\centering
\begin{tikzpicture}[x=0.6pt,y=0.6pt,yscale=-1,xscale=1]

\draw (0,0) .. controls (100,-30) .. (300,-20);
\draw (0,0) .. controls (-100,-30) .. (-200,-40);

\draw[ultra thick]  (0,0) .. controls (80,-50)  .. (200,-150);

\draw[dashed] (-200,-40) .. controls (0,-180) .. (200,-150);
\draw[dashed] (300,-20) .. controls (210,-40) .. (200,-150);

\draw[->,ultra thick] (79,-50)--(81,-51);
\draw[->,thick] (-99,-28)--(-100,-28);
\draw[->,thick] (98,-25)--(100,-25);

\filldraw[black] (300,-20) circle (3pt) node[anchor=north]{};
\filldraw[black] (-200,-40) circle (3pt) node[anchor=north]{};
\filldraw[blue] (0,0) circle (4pt) node[anchor=north]{};
\filldraw[red] (200,-150) circle (4pt) node[anchor=north]{};
\draw (-20,10) node [anchor=north west][inner sep=0.75pt]  [font=\large]  {Ising fixed point};
\draw (200,-170) node [anchor=north west][inner sep=0.75pt]  [font=\large]  {Yang-Lee fixed point};
\draw (150,-50) node [anchor=north west][inner sep=0.75pt]  [font=\large]  {$i\sigma$};
\draw (-80,-40) node [anchor=north west][inner sep=0.75pt]  [font=\large]  {$\epsilon$};
\draw (-230,-50) node [anchor=north west][inner sep=0.75pt]  [font=\large]  {F};
\draw[color=red!60, dashed](200,-150) circle (40);
\end{tikzpicture}
\caption{Scheme of RG flows for imaginary value of the magnetic field in the Ising field theory. A combination of the thermal deformation and a purely imaginary magnetic deformation flows to the Yang-Lee fixed point. The latter is controlled by the minimal model $\mathcal M(2,5)$. The point F corresponding to pure thermal perturbation is the theory of a single massive Majorana fermion.}
\label{RGflowIsing}
\end{figure}
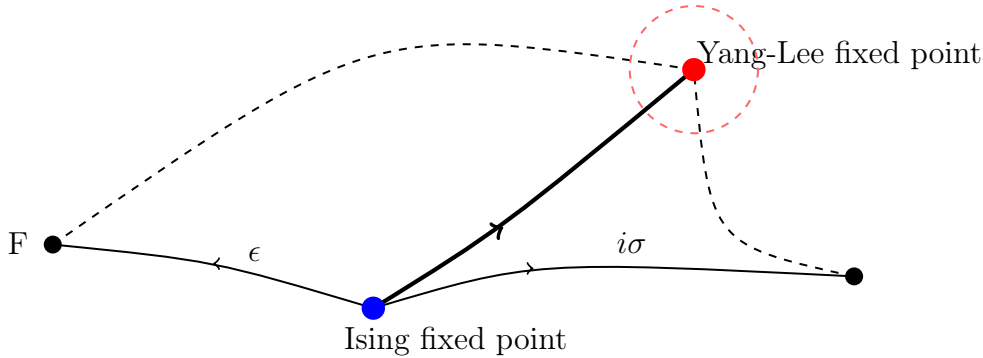

In the case of the Lagrangian \eqref{IsingFermionLagr} the Hamiltonian satisfies the property $i)$. Indeed the explicit action of the $\op P \op T$ transformation is given by 
\begin{equation}
x \to - x \ , \hspace{1 cm} i \to -i \ , \hspace{1 cm} \psi \to i \psi \ , \hspace{1 cm} \overline \psi \to i \overline \psi  \ , \hspace{1 cm}\sigma \to - \sigma \ .
\end{equation}
The property $ii)$ is much less trivial to verify. Numerical computations (e.g. \cite{Fonseca:2001dc,Xu:2022mmw}) show that for the imaginary magnetic field below a certain critical value $h<h_c$  the energy levels are real, while for $h>h_c$ they become complex, consistent with a $\op P \op T$-symmetric / spontaneously broken $\op P \op T$ phase, respectively. The conclusion is that the Yang-Lee fixed point separates a $\op P\op T$-symmetric phase from a $\op P \op T$-breaking phase. We summarise the different RG flows in Fig.~\ref{RGflowIsing}.

The $\op P \op T$-symmetry of the system is important also for the properties of the RG flow. Indeed the original $c$-theorem, proposed and proved by Zamolodchikov \cite{Zamolodchikov:1986gt}, can be extended for non-unitary RG flows if the $\op P \op T$ symmetry is preserved \cite{Castro-Alvaredo:2017udm}, implying the existence of a monotonically decreasing function interpolating between the ultraviolet and infrared effective central charges, where the effective central charge is defined as  $c_\text{eff} = c- 24 \Delta_\text{min}$ with $\Delta_\text{min}$ being the minimum value of conformal dimension in the given CFT, which is negative for non-unitary models and zero for unitary ones. The application of this theorem is justified here since the Lee-Yang critical point can be approached from fully inside the $\op P\op T$-symmetric phase. For the Ising model in imaginary magnetic field, the above generalisation of the $c$-theorem then provides a rigorous bound on the effective central charge of the infrared theory: 
\begin{equation}
c_\text{eff}^\text{IR} < c_\text{eff}^\text{UV} = \frac{1}{2} \ .
\end{equation}
From the expression for the effective central charge minimal model $\mathcal M(p,q)$
\begin{equation}\label{ceffformula}
	c_\text{eff} = 1-\frac{6}{pq} \ ,
\end{equation}
we obtain the condition $pq<12$ which is only satisfied by the minimal model $\mathcal M(2,5)$, consistently with Cardy's prediction.

\subsection{TCSA phenomenology for the Yang-Lee edge singularity in the Ising model}

Numerical studies on the Yang-Lee edge singularities have already been performed in the literature. In particular, lattice studies were the first to confirm that the critical point is controlled by the minimal model $\mathcal M(2,5)$ \cite{Itzykson:1986pk}, followed more recently by field theoretical ones \cite{Fonseca:2001dc,Xu:2022mmw}. The latter works applied a Hamiltonian truncation called \textit{Truncated free fermionic space approach} (TFFSA), which exploits the fact that for $h=0$ the model can be described using a free massive Majorana fermion. TFFSA uses the corresponding Hilbert space as a computational basis, utilising the fact that the exact finite volume matrix elements of the order field $\sigma$ are known \cite{2001hep.th....7117B}. Unfortunately this approach cannot be generalised to higher unitary minimal models $\mathcal M_{p,p+1}$  since the exact expression of finite volume matrix elements of relevant scaling fields in the thermally deformed model are not known in general. Therefore it is necessary to resort to the TCSA formulation which uses the CFT to provide the conformal basis. In order to benchmark this method, here we present it for the case of critical Ising model to demonstrate that it can reproduce the evidence for the universality class of the Yang-Lee edge singularity. 

Therefore let us consider the Hamiltonian 
\begin{equation}\label{IsingHamiltonian}
 \op  H  = \op H_{\mathcal M(3,4)}+ \frac{m}{2 \pi} \int \epsilon(x,y) \ \de x + i h \int \sigma(x,y) \ \de x  \ ,
\end{equation}
where $\op H_{\mathcal M(3,4)}$ is the conformal Hamiltonian of the minimal model $\mathcal M(3,4)$. The mass $m$ of the free Majorana fermion obtained in the limit $h=0$ provides a set of units, and the only physical parameter of the theory is the dimensionless ratio \begin{equation}
	\xi = \frac{h}{|m|^{15/8}} \ .
\end{equation}

\begin{figure}[htb]
\begin{subfigure}[t]{.45\textwidth}
  \centering
  \includegraphics[width=0.98\textwidth]{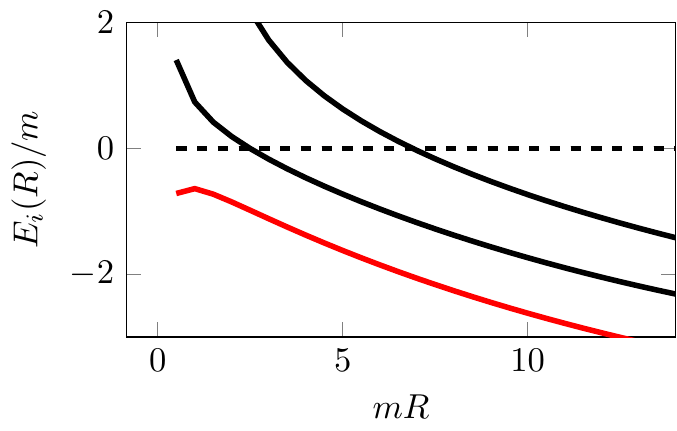}
  \caption{First three energy levels in the theory for $\xi = 0.1<\xi_c$ at level cut-off $N_\text{max}=18$. The solid lines represent the real part of the energy while the dashed lines represent the imaginary part, with the ground state highlighted in red.}
  \label{fig:Ising01}
\end{subfigure}%
\hfill
\begin{subfigure}[t]{.45\textwidth}
  \centering
  \includegraphics[width=0.98\linewidth]{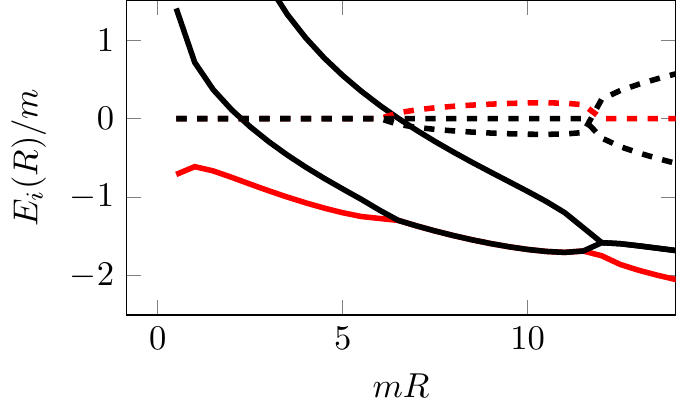}
  \caption{First three energy levels in the theory for $\xi = 0.2>\xi_c$ at level cut-off $N_\text{max}=18$. The solid lines represent the real part of the energy while the dashed lines represent the imaginary part, with the ground state highlighted in red.}
  \label{fig:Ising02}
\end{subfigure}
\label{fig:Ising_LY_transition}
\caption{Comparing the finite volume spectrum in the $\op P\op T$-symmetric phase $\xi<\xi_c$ to the one in the $\op P\op T$-breaking phase $\xi>\xi_c$ in the Ising model.}
\end{figure}

The critical point in the RG flow can be identified as the value $\xi_c$ for which the ground state and the second excited state meet to form a complex conjugate pair. In the exact spectrum this happens first at $R=\infty$, with the branching point moving down in the volume as $\xi$ is increased further. At the critical point, the volume dependent mass gap $M(R)\simeq  E_1(R)-E_0(R)$ goes to zero asymptotically as $R\to\infty$ corresponding to a divergent  correlation length ($1/M$) explodes. The expected phenomenology is the following: 
 \begin{itemize}
\item In the region $\xi < \xi_c$ the ground state is real (\ref{fig:Ising01}) and the system is in the $\op P\op T$-symmetric phase;
\item For $\xi > \xi_c$ the ground state and the first excited state form a complex conjugate pair in large volumes (\ref{fig:Ising02}) and the system is in the $\op P\op T$-breaking phase.
\end{itemize}
The truncation inherent in TCSA results in a running coupling dependent on the cut-off \cite{2011arXiv1106.2448G}. As a result, the actual branching of the two levels happens at a cut-off dependent value $\xi_c(\Lambda)$ and the exact phenomenology can only be reproduced in the limit of infinite cut-off. In addition, truncation effects also result in the branching point appearing in a finite volume $R$ which can be increased by raising the truncation level and by carefully tuning the parameter $\xi$, but only within some limits due to finite computing resources such as memory, numerical accuracy and CPU time.

Tuning carefully the parameter $\xi$, for the cut-off $N_\text{max}=18$ one arrives at the following critical value separating the two phases
\begin{equation}\label{Isingcriticalvalue}
    \xi_c = 0.1905\dots\ .
\end{equation}
We note that when this value slightly deviates from that reconstructed from the results obtained by TFFSA \cite{Fonseca:2001dc,Xu:2022mmw}. This is eventually fully expected, since the critical value $\xi_c$ depends on the cut-off, and the TCSA and TFFSA use different cut-off procedures due to differing choices of the computational basis. Concerning the cut-off dependence we note that due to the dimension of the energy operator, the Ising TCSA is eventually logarithmically divergent \cite{2011arXiv1106.2448G}. While the divergence can be offset by a counter term proportional to the identity and therefore cancels from relative energy levels, the sensitivity to the UV cut-off is enhanced to $\Lambda^{-1}$ following from the level dependent subleading corrections. Since the TFFSA treats the thermal perturbation in an exact analytic way exploiting the free massive Majorana fermion description, it has much better convergence properties in terms of the cut-off.

Once the critical value is determined, one can turn to a detailed study of the energy spectrum to find the physical window where the scaling function $(E_1(R)-E_0(R))R$ is approximately constant. With our choice of the cut-off it happens for $mR\lesssim 6$, where we can extract the scaling functions $C_i(R) = (E_i-E_0) R/(4 \pi)$. Under the assumption of the IR fixed point being described by the minimal model $\mathcal M(2,5)$, the ground state  must flow to a conformal state corresponding to a primary field $\varphi$ with conformal dimensions $(-1/5,-1/5)$, the first excited state is expected to flow to the state corresponding to the identity field with conformal dimensions $(0,0)$, while the second level to a descendent state corresponding to $\op L_{-1}\op{\overline{L}}_{-1}\varphi$ with conformal dimensions $(4/5,4/5)$, resulting in the following predictions for the large volume asymptotics of the scaling functions:
\begin{equation}\label{M25Predictions}
    C_1(R=\infty) = \frac{1}{5} \, , \hspace{1 cm} C_2(R=\infty) = 1 \, .
  \end{equation}

\begin{figure}[htb]
\begin{subfigure}[t]{.45\textwidth}
  \centering
  \includegraphics[width=0.98\textwidth]{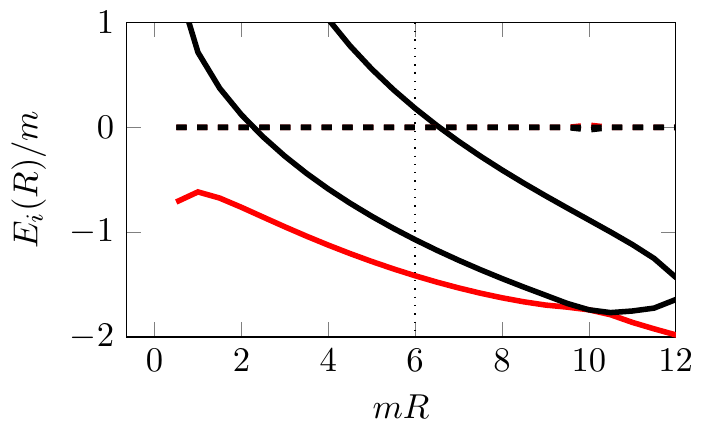}
  \caption{The first three energy levels, with the solid lines showing the real parts of the energies, while the dashed lines depict the imaginary parts, with the ground state highlighted in red. The dotted vertical line indicates the end of the physical window.}
  \label{fig:IsingRaw}
\end{subfigure}%
\hfill
\begin{subfigure}[t]{.45\textwidth}
  \centering
  \includegraphics[width=0.98\linewidth]{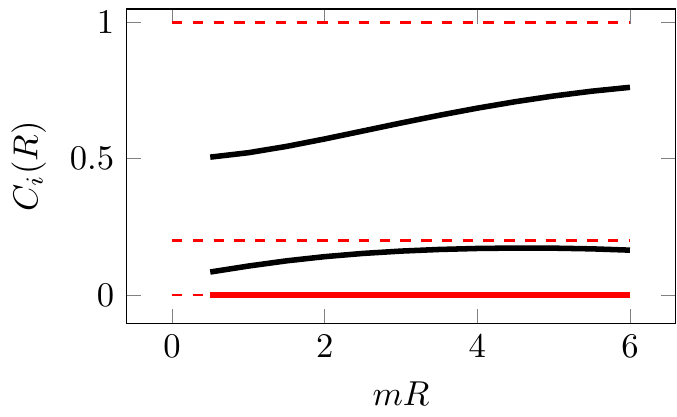}
  \caption{The scaling functions $C_i(R) = R/(4 \pi)(E_i-E_0)$ corresponding to the first three energy levels for $mR\leq 6$. The solid lines are the TCSA results, while the dashed lines display the asymptotic values expected from $\mathcal M(2,5)$.}
  \label{fig:IsingC}
\end{subfigure}
\caption{The spectrum of the Ising model in an imaginary magnetic field with the value $\xi=\xi_c$ in equation \eqref{Isingcriticalvalue}, at level cut-off $N_\text{max}=18$. }\label{fig:IsingLY_TCSA}
\end{figure}
As shown in Fig. \ref{fig:IsingLY_TCSA}, the numerical results obtained from the TCSA are fully consistent with the predictions for the fixed point described by the minimal model $\mathcal M(2,5)$, and very similar to those obtained using the TFFSA \cite{Fonseca:2001dc,Xu:2022mmw}. Note in particular that the dimension of the single nontrivial primary field is reproduced quite accurately, which is going to play an important role in the tricritical case to which we now turn.

\section{Tricritical Ising field theory in imaginary magnetic field}\label{sec:TIM_LY}

\subsection{The RG flow triggered by imaginary magnetic fields}\label{subsec:RG_imag_magnetic}
Now we turn to considering the RG flow starting from the minimal model $\mathcal M(4,5)$ triggered by an imaginary magnetic field. Even though the distribution of the Yang-Lee zeros is not known analytically in lattice models in this universality class such as e.g. the Blume-Capel model \cite{BC1,BC2,BC3,BC4}, the presence of a phase transition can be understood in analogy to the Ising case as a divergence in the density of Yang-Lee zeros, and indeed this scenario has already been numerically tested on the lattice \cite{vonGehlen:1994rp}. Here we consider the field theory counterpart of this phenomenon in parallel to the critical Ising case. 
\begin{table}[b]
\begin{center}
\begin{tabular}{ c c c c}
\hline 
Primary & Weights & LG field & name \\ 
$\op 1$ & (0,0) &  1 & Identity \\
$\sigma$ & (3/80,3/80) & $\varphi$ & magnetisation \\
$\epsilon$ & (1/10,1/10) & $:\varphi^2:$ & energy \\
$\sigma'$ & (7/16,7/16) & $:\varphi^3:$ & subleading magnetisation \\
$t$ & (3/5,3/5) & $:\varphi^4:$ & vacancy density \\
$\epsilon''$ & (3/2,3/2) & $:\varphi^6:$ & (irrelevant) \\
 \hline 
\end{tabular}
\caption{Primary fields in the Tricritical Ising model $\mathcal M(4,5)$, listing their conformal weights and their identification in the Landau-Ginzburg description.}
\label{TIMContent}
\end{center}
\end{table}

The primary fields of the minimal model $\mathcal M(4,5)$ are listed in Table~\ref{TIMContent} together with their weights and the Landau-Ginzburg identification as normal ordered powers of the order parameter (chosen to be the magnetic field). The scaling region of the tricritical Ising model is spanned by the four relevant deformations of the minimal model $\mathcal M(4,5)$, and it was previously examined for the case when all the coupling constants real in Refs. \cite{Lassig:1990xy, Lepori:2008et,Lencses:2021igo}, while the complex analytic properties of the free energy were studied in Ref. \cite{Mossa:2007fx}.

Note that half of the relevant fields is are even under the $\mathbb{Z}_2$ symmetry of the spin model (corresponding to even powers of the Landau-Ginzburg field $\varphi$) while the other half is odd (corresponding to odd powers $\varphi$). The Landau-Ginzburg action has the form
\begin{equation}\label{TricriticalLandauGinzburg}
\begin{split}
\mathcal L_\text{TIM}= \frac{1}{2}\partial_\mu \varphi \partial^\mu \varphi + g_1 \varphi + g_2 \varphi^2+ g_3 \varphi^3+ g_4 \varphi^4 + \varphi^6 \ .
\end{split}
\end{equation}
When considering the phase diagram, the odd parameters $g_1$ and $g_3$ corresponding to explicit symmetry breaking are switched off and therefore the phases are parameterised using only the even couplings $g_2$ and $g_4$. 

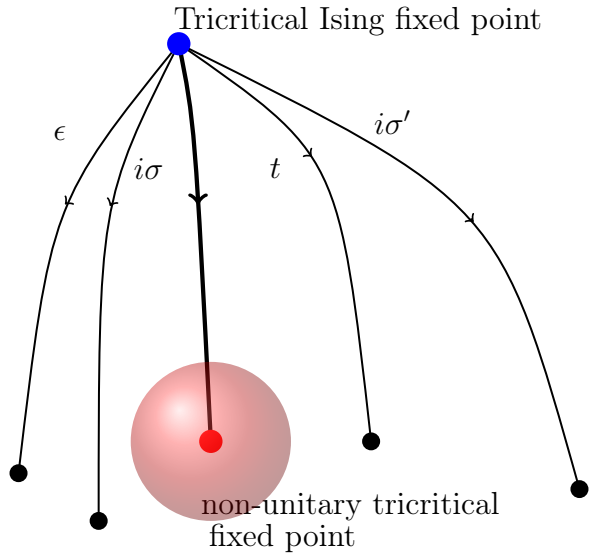
\begin{figure}[t]
\centering
\begin{tikzpicture}[x=0.6pt,y=0.6pt,yscale=-1,xscale=1]
\draw (0,0) .. controls (-80, 100) .. (-100, 270) ;
\draw (0,0) .. controls (200, 100) .. (250, 280) ;
\draw (0,0) .. controls (-50, 100) .. (-50, 300) ;
\draw (0,0) .. controls (100, 70) .. (120, 250) ;
\draw[ultra thick]  (0,0) .. controls (10,50)  .. (20,250);

\draw[->,thick] (-70,100)--(-71,101);
\draw[->,thick] (-42,100)--(-43,102);
\draw[->,thick] (182,110)--(184,112);
\draw[->,thick] (82,70)--(83,71);
\draw[->,ultra thick] (12,100)--(12,101);

\filldraw[black] (-100,270) circle (3pt) node[anchor=north]{};
\filldraw[blue] (0,0) circle (4pt) node[anchor=north]{};
\filldraw[black] (250,280) circle (3pt) node[anchor=north]{};
\filldraw[black] (-50,300) circle (3pt) node[anchor=north]{};
\filldraw[black] (120,250) circle (3pt) node[anchor=north]{};
\filldraw[red] (20,250) circle (4pt) node[anchor=north]{};

\draw (-5,-25) node [anchor=north west][inner sep=0.75pt]  [font=\large]  {Tricritical Ising fixed point};
\draw (12,280) node [anchor=north west][inner sep=0.75pt]  [font=\large]  {non-unitary tricritical};
\draw (17,300) node [anchor=north west][inner sep=0.75pt]  [font=\large]  {fixed point};
\draw (-80,50) node [anchor=north west][inner sep=0.75pt]  [font=\large]  {$\epsilon$};
\draw (-30,70) node [anchor=north west][inner sep=0.75pt]  [font=\large]  {$i\sigma$};
\draw (120,40) node [anchor=north west][inner sep=0.75pt]  [font=\large]  {$i\sigma'$};
\draw (55,70) node [anchor=north west][inner sep=0.75pt]  [font=\large]  {$t$};
\shade[ball color = red!, opacity = 0.4] (20,250) circle (50);
\end{tikzpicture}
\caption{Scheme of RG flows for imaginary value of the (physical) magnetic field. A combination of the purely imaginary leading and subleading magnetic deformations, supplemented by thermal and vacancy deformations flows to the non-unitary tricritical fixed point.}
\label{RGflowTricritical}
\end{figure}

A crucial observation is that the physical magnetic field which couples to the spins in the lattice, scales to a nontrivial combination involving all the odd fields in the QFT. In the case of $\mathcal M(4,5)$ there are two relevant odd fields: the magnetisation $\sigma$ and the subleading magnetisation $\sigma'$. To reach a possible non-unitary tricritical point it is necessary to include both of as deformations with imaginary couplings. Furthermore, in analogy with the Ising case, even relevant fields should also be included as they are generated along the renormalisation flow, however their couplings are real as they arise from operator products containing an even number of the odd fields. The resulting Hamiltonian then has the form 
\begin{equation}\label{TricriticalHamiltonian}
\begin{split}
\op H = & \op H_{\mathcal M(4,5)}+ \mu \int \epsilon(x,y) \ \de x+ i h \int \sigma(x,y)\ \de x  +\\ & +i h' \int \sigma'(x,y) \ \de x  + v \int t(x,y) \ \de x  \ .
\end{split}
\end{equation}
and similarly to the Ising model it is $\op P \op T$ symmetric i.e. $[\op H,\op P \op T] = 0$, with the related discussion carrying without any essential change. The Yang-Lee edge singularity is located on the boundary between $\op P \op T$-symmetric and $\op P \op T$-breaking phases, with the flow reaching it from inside the $\op P \op T$-symmetric phase. As a result, the generalised $c$-theorem \cite{Castro-Alvaredo:2017udm} gives an upper bound on the effective central charge of the infrared fixed point: 
\begin{equation}
	c_\text{eff}^\text{IR}< c_\text{eff}^\text{UV} = \frac{7}{10} \ , 
\end{equation}
which, using the parameterisation (\ref{centralcharge}) of the central charges, means that $pq < 20$. This bound is only satisfied by the minimal models $\mathcal M(2,5)$, $\mathcal M(2,7)$, $\mathcal M(2,9)$ and $\mathcal M(3,5)$. Note that in the presence of imaginary couplings for the odd fields as written in Eq. \eqref{TricriticalHamiltonian} all terms are invariant under the action of the $\op P \op T$ symmetry and therefore appear on equal footing when considering the Yang-Lee edge singularity. The various possible RG flows are depicted in Fig.~\ref{RGflowTricritical}.

Let us attempt to adapt Fisher's argument for the tricritical case to determine the Landau-Ginzburg description of the infrared fixed point. In terms of Landau-Ginzburg Lagrangian density \eqref{TricriticalLandauGinzburg} the tricritical point is located at $g_1 = g_2 = g_3 = g_4 = 0$. Shifting the order parameter with an imaginary constant term $\varphi \to \varphi+ i \varphi_0$ and fixing the coupling constants and $\varphi_0$ to preserve tricriticality, i.e. setting  couplings of $\varphi^2$, $\varphi^3$ and $\varphi^4$ to zero, it results in the theory 
\begin{equation}\label{varphi^5}
\mathcal L_{NHT}= \frac{1}{2}\partial_\mu \varphi \partial^\mu \varphi+(h-i h_0)\varphi + i \gamma \varphi^5+\dots \ .
\end{equation}
The above Lagrangian implies a field equation 
\begin{equation}
    \partial^\mu \partial_\mu\varphi \propto \varphi^4
\end{equation}
which implies the existence of three independent relevant scaling fields $\varphi$, $\varphi^2$ and $\varphi^3$ (besides the identity). The only minimal models with this property are $\mathcal M(2,9)$ and $\mathcal M(3,5)$, however, their operator algebra is inconsistent with the above equation of motion, as shown in appendix \ref{sec:AppendixA}).

To understand what goes wrong here, note that in order to obtain the Lagrangian in \eqref{varphi^5} the parameters $\varphi_0, g_1, g_2, g_3$ and $g_4$ were treated as if they were independent. However, in light of the RG flow determined by von Gehlen on the lattice \cite{vonGehlen:1994rp}, this is not a correct assumption. In order to reach the critical surface (which in this case is expected to be a line of ordinary critical points) it is necessary to tune the mass gap to zero, i.e. to make the ground state and the first excited state meet at $R\to\infty$. The ends of this critical line are then expected to correspond to a different universality class corresponding to the \textit{non-unitary tricritical point}, where the ground state meets simultaneously with both the first and the second excited states. It is then clear that to stay on the (one-dimensional) critical line the couplings  $\varphi_0, g_1, g_2, g_3$ and $g_4$ cannot be varied independently, which prevents a straightforward extension of Fisher's argument to this case. Therefore the correct form of the Landau-Ginzburg potential is not obvious at this stage.

Nevertheless, we can surmise the number of independent relevant fields from the RG perspective. In contrast to the unitary case of multicritical Ising points discussed above (where half of the relevant fields were order parameters which were odd under the $\mathbb{Z}_2$ symmetry), there are no such fields in the $PT$-symmetric non-unitary multicritical Lee-Yang points and deformations by all relevant fields preserve $PT$ invariance\footnote{Here we remark that preservation of $PT$ at the Hamiltonian level prescribes the couplings of $PT$-even/odd perturbing fields to be purely real/imaginary, respectively. For example, the Lee-Yang model must be perturbed by an imaginary multiple of its only nontrivial primary, when the latter is normalised to have a short distance singularity with a coefficient of unity \cite{Yurov:1989yu}.} in a domain of the couplings which has a boundary set by spontaneous breaking of $PT$ symmetry. As a result, we expect that the $(m+1)$-th multicritical Lee-Yang points form a codimension-one boundary of the $m$-th multicritical Lee-Yang points, and so it has exactly one more relevant operator. This leads to the expectation that while the ordinary Lee-Yang critical point has a single non-trivial relevant field, the tricritical Lee-Yang point has exactly two. This argument singles out the minimal model $\mathcal M(2,7)$ as the main candidate for the universality class of the tricritical Yang-Lee edge singularity. The minimal model $\mathcal M(2,7)$ contains three primary fields $\phi$, $\phi'$, and the identity $\op 1$ with conformal weights $(-3/7,-3/7)$, $(-2/7,-2/7)$  and $(0,0)$, respectively. 

Based on the considerations above, the Lee-Yang critical points in the model $\mathcal M(4,5)$ are expected to form a critical line which is controlled by the minimal model $\mathcal M(2,5)$, the tricritical endpoints of which are governed by the minimal model $\mathcal M(2,7)$ which thus describes the Lee-Yang tricriticality. Such a  picture appears to be reasonable also because, for real couplings, the critical line in the scaling region of $\mathcal M(4,5)$ is controlled by the Ising universality class $\mathcal M(3,4)$, which flows to $\mathcal M(2,5)$ under the influence of an imaginary magnetic field. Furthermore, this proposal turns out to be also in agreement with the numerical results obtained on the  lattice \cite{vonGehlen:1994rp}.

\subsection{TCSA phenomenology in the $v = 0$ section}\label{Phenomenology}
\begin{figure}[htb]
\begin{subfigure}[t]{.45\textwidth}
  \centering 
  \includegraphics[width=0.98\textwidth]{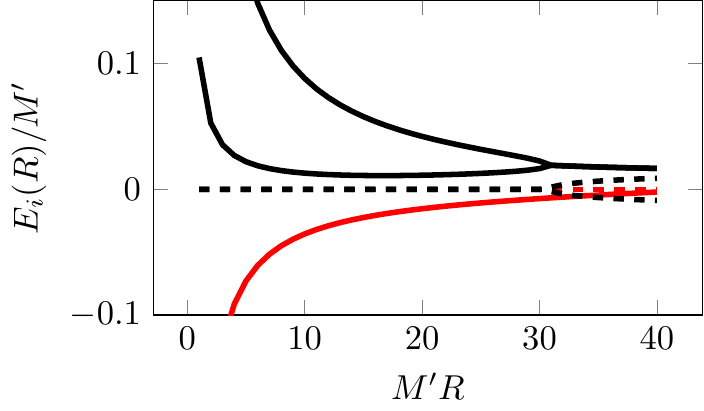}
  \caption{First three energy levels in the theory for $(\zeta',\zeta ) =(3.17,0)$. The solid lines represent the real part of the energy while the dashed lines represent the imaginary part. The ground state is highlighted in red.}
   \label{fig:zeta01}
\end{subfigure}%
\hfill
\begin{subfigure}[t]{.45\textwidth}
 \centering
  \includegraphics[width=0.98\linewidth]{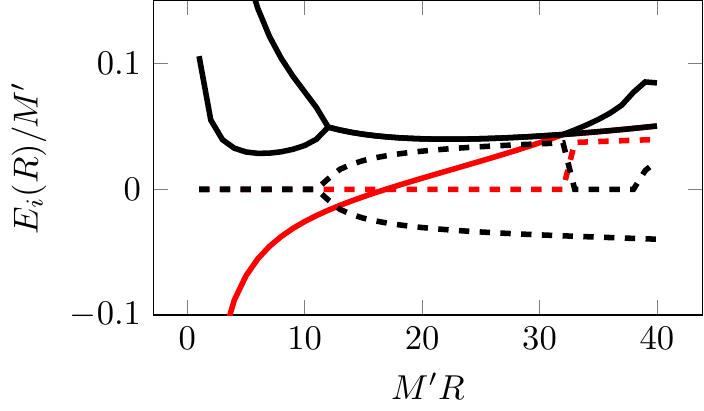}
  \caption{First three energy levels in the theory for $(\zeta',\zeta ) =(7.93,0)$. The solid lines represent the real part of the energy while the dashed lines represent the imaginary part. The ground state is highlighted in red.}
  \label{fig:zeta02}
\end{subfigure}
\caption{The typical behaviours in the tricritical Ising model observed in the $\zeta = 0$, at a level cut-off $N_\text{max}=10$.}
\end{figure}

\begin{figure}[htb]
\begin{subfigure}[t]{.45\textwidth}
  \centering 
  \includegraphics[width=0.98\textwidth]{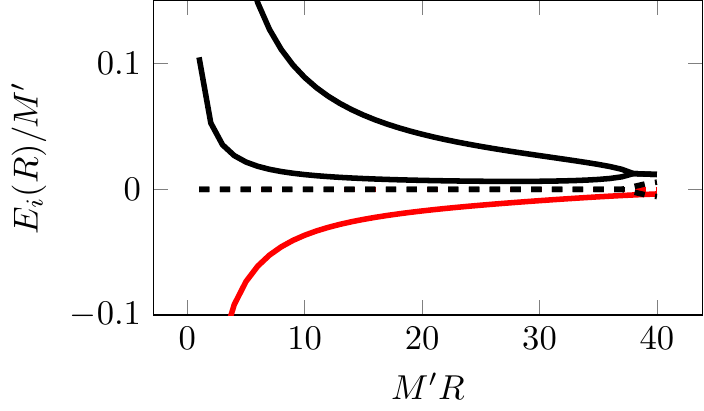}
  \caption{First three energy levels in the theory for $(\zeta',\zeta ) =(0,3.67)$. The solid lines represent the real part of the energy while the dashed lines represent the imaginary part. The ground state is highlighted in red.}
   \label{fig:zetap01}
\end{subfigure}%
\hfill
\begin{subfigure}[t]{.45\textwidth}
 \centering
  \includegraphics[width=0.98\linewidth]{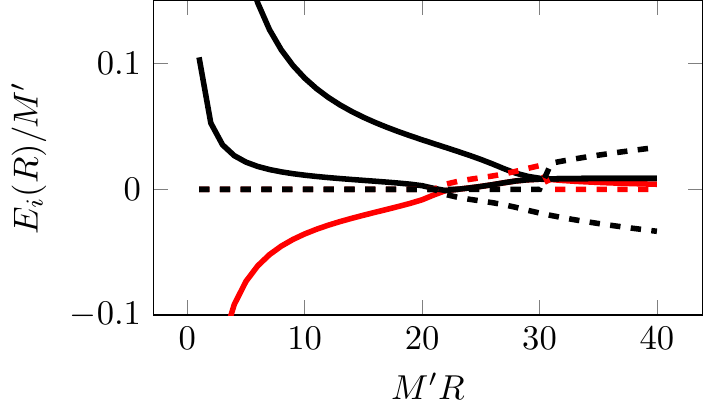}
  \caption{First three energy levels in the theory for $(\zeta',\zeta ) =(0,9.17)$. The solid lines represent the real part of the energy while the dashed lines represent the imaginary part. The ground state is highlighted in red.}
  \label{fig:zetap02}
\end{subfigure}
\caption{The typical behaviours in the tricritical Ising model observed in the $\zeta' = 0$, at a level cut-off $N_\text{max}=10$.}
\end{figure}

To apply TCSA we put the Hamiltonian \eqref{TricriticalHamiltonian} on the cylinder. Just like for the Ising model, the coupling of the energy operator can be used to set the units. In fact, for $h=h'=v=0$ the model \eqref{TricriticalHamiltonian} is integrable with the scattering described by the minimal $E_7$ factorised $S$ matrix~\cite{MC,FZ}. For this case the exact expression of the mass gap in terms of the coupling $\mu$ is known \cite{Fateev:1993av}, and it is possible to use it to set the units in analogy with the Ising case. However, it turns out that choosing a different unit is often convenient for visualisation of finite volume the spectrum along the RG flow, and for the purposes of this subsection we choose  a mass scale $M'$ by setting 
\begin{equation}
    \mu = \frac{10^{-4}}{2 \pi} {M'}^{9/5}
\end{equation}
With the above choice the model has only two dimensionless parameters
\begin{equation}
\zeta = \frac{h}{{M'}^{77/72}} \ , \hspace{1 cm} \zeta' = \frac{h'}{{M'}^{5/8}} \ ,
\end{equation} 
We probe the two-dimensional space parameterised by $\zeta$ and $\zeta'$ by varying $\zeta$ for different values of $\zeta'$. Due to the odd nature of $\sigma $  and $\sigma'$ only the relative sign of $h$ matters, therefore we restrict $h$ (and consequently $\zeta$) to positive values, explicitly, since $\sigma \to -\sigma$ and $\sigma' \to - \sigma'$ is a symmetry of the Hamiltonian in equation \eqref{TricriticalHamiltonian}, then the phase diagram for the $v = 0$ plane is expected to be symmetric with respect to the origin.
Let us first discuss the two axes: the symmetry with respect to the origin implies that it is sufficient to consider only positive values of the couplings. On the $\zeta  = 0$ axis the first excited state always meets the second excited states before the ground state, as it is shown in figures \ref{fig:zeta01} and \ref{fig:zeta02}.
On the $\zeta' = 0$ axis we the first excited state meets the second excited state before the ground state for small values of $\zeta$ (figure \ref{fig:zetap01}), i.e. near the $E7$ integrable model in the origin; for large values of the coupling $\zeta$ the ground state meets the first excited state first (figure \ref{fig:zetap02}). This suggest that there is a point in the $\zeta' = 0$ axis in which the ground state meets the first excited state and the second excited state in a single point for a finite value of the dimensionless volume $M' R$. 

\begin{figure}[htb]
\begin{subfigure}[t]{.45\textwidth}
  \centering
  \includegraphics[width=0.98\textwidth]{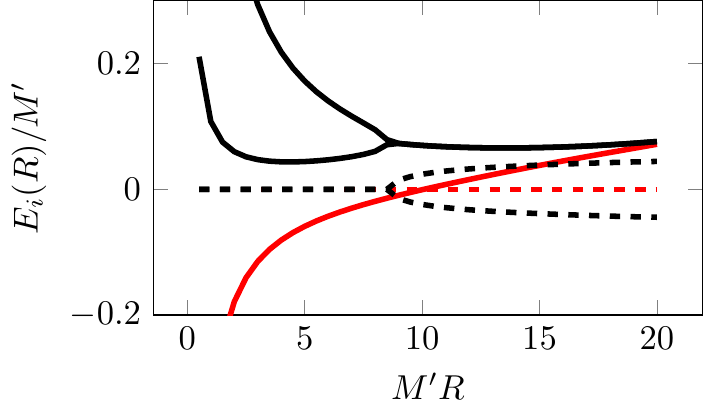}
  \caption{First three energy levels in the theory for $(\zeta',\zeta ) =(-31.62,1.90)$. The solid lines represent the real part of the energy while the dashed lines represent the imaginary part. The ground state is highlighted in red.}
   \label{fig:n0a01}
\end{subfigure}%
\hfill
\begin{subfigure}[t]{.45\textwidth}
 \centering
  \includegraphics[width=0.98\linewidth]{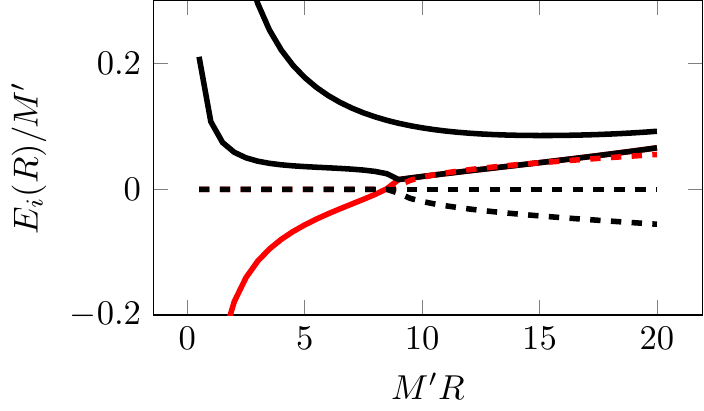}
  \caption{First three energy levels in the theory for $(\zeta',\zeta ) =(-31.62,5.69)$. The solid lines represent the real part of the energy while the dashed lines represent the imaginary part. The ground state is highlighted in red.}
  \label{fig:n0a03}
\end{subfigure}
\caption{The typical behaviours in the tricritical Ising model observed for a large negative value of $\zeta'$, at a level cut-off $N_\text{max}=10$.}
\end{figure}

\begin{figure}[htb]
\begin{subfigure}[t]{.45\textwidth}
  \centering
  \includegraphics[width=0.98\textwidth]{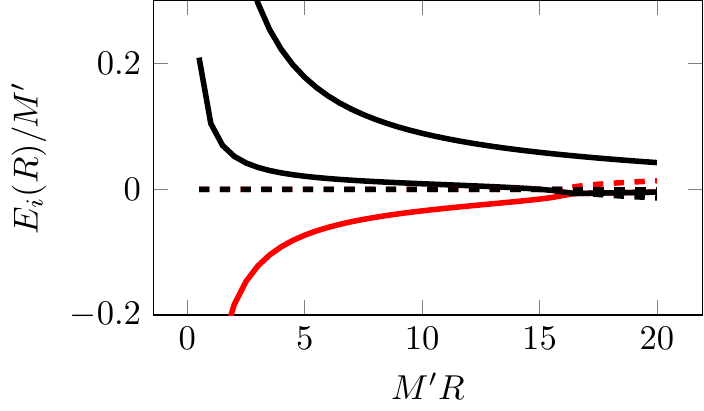}
  \caption{First three energy levels in the theory for $(\zeta',\zeta ) =(-3.16,18.60)$. The solid lines represent the real part of the energy while the dashed lines represent the imaginary part. The ground state is highlighted in red. }
  \label{fig:n00a001}
\end{subfigure}%
\hfill
\begin{subfigure}[t]{.45\textwidth}
  \centering
  \includegraphics[width=0.98\linewidth]{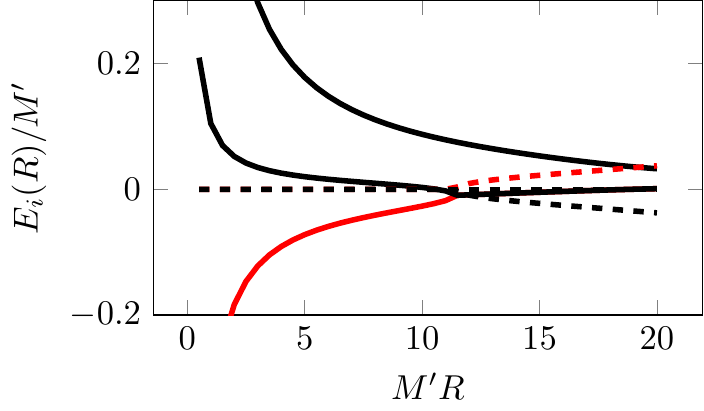}
  \caption{First three energy levels in the theory for $(\zeta',\zeta ) =(-3.16,37.91)$. The solid lines represent the real part of the energy while the dashed lines represent the imaginary part. The ground state is highlighted in red.}
  \label{fig:n00a02}
\end{subfigure}
\caption{The typical behaviours observed in the tricritical Ising model observed for a less negative value of $\zeta'$, at a level cut-off $N_\text{max}=10$.}
\end{figure}

Let us now discuss the behaviours of the energies away from the two axes. For large negative values of $\zeta'$ and for small $\zeta$, the first and the second exited states meet to form a complex conjugate pair for a finite value of the volume (Fig. \ref{fig:n0a01}), however increasing $\zeta$ it is possible to change this picture into the ground state and the first excited state meeting first (Fig. \ref{fig:n0a03}). 
Making $\zeta'$ less negative, the picture changes in such a way that for small and large values of $\zeta$ (Fig. \ref{fig:n00a001} and \ref{fig:n00a02}) the ground state  meets the first excited state in a complex conjugate pair (Fig. \ref{fig:n00a02}) before the second and the first excited states could meet.
It is natural to expect that keeping $\zeta'<0$ and varying carefully the two parameter $\zeta$ and $\zeta'$, there should exist a value $(\zeta'_{c,1},\zeta_{c,1})$ of the couplings when the lowest three energy levels meet simultaneously. This is the point the $v=0$ plane which is closest to one of the endpoints of the critical line that corresponds to the non-unitary tricritical Lee-Yang edge singularity. To eventually hit the tricritical point requires switching on $v\neq 0$ and tuning it together with $\zeta$ and $\zeta'$ to put the triple meeting point towards (ideally) infinite volume.

\begin{figure}[htb]
\begin{subfigure}[t]{.45\textwidth}
  \centering
  \includegraphics[width=0.98\textwidth]{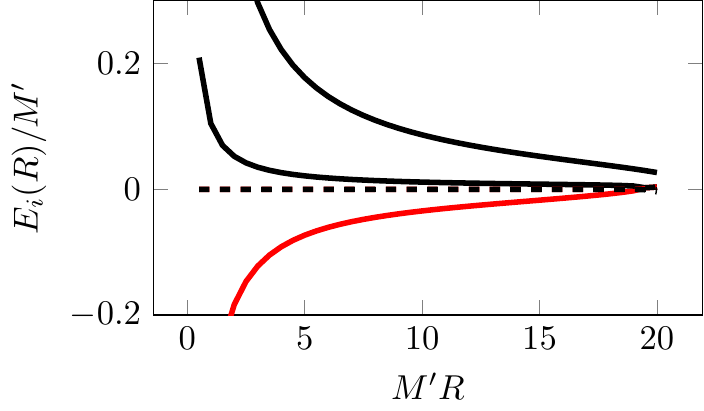}
  \caption{First three energy levels in the theory for $(\zeta',\zeta ) =(3.16,18.60)$. The solid lines represent the real part of the energy while the dashed lines represent the imaginary part. The ground state is highlighted in red.}
  \label{fig:00a001}
\end{subfigure}%
\hfill
\begin{subfigure}[t]{.45\textwidth}
  \centering
  \includegraphics[width=0.98\linewidth]{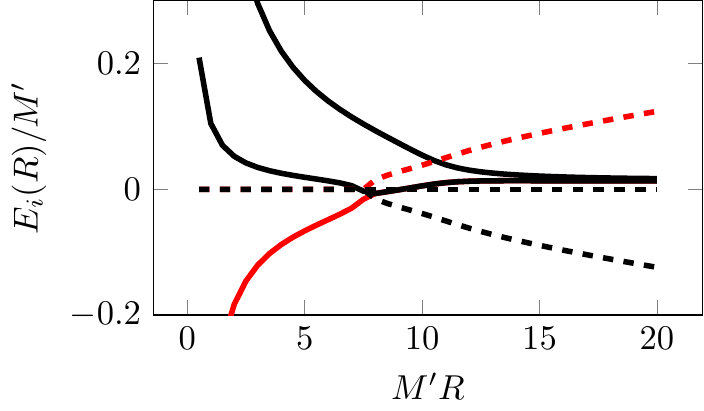}
  \caption{First three energy levels in the theory for $(\zeta',\zeta ) =(3.16,94.79)$ . The solid lines represent the real part of the energy while the dashed lines represent the imaginary part. The ground state is highlighted in red.}
  \label{fig:00a05} 
\end{subfigure}
\caption{The typical behaviours observed in the tricritical Ising model observed for a smaller positive value of $\zeta'$, at a level cut-off $N_\text{max}=10$.}
\end{figure}

\begin{figure}[htb]
\begin{subfigure}[t]{.45\textwidth}
  \centering
  \includegraphics[width=0.98\textwidth]{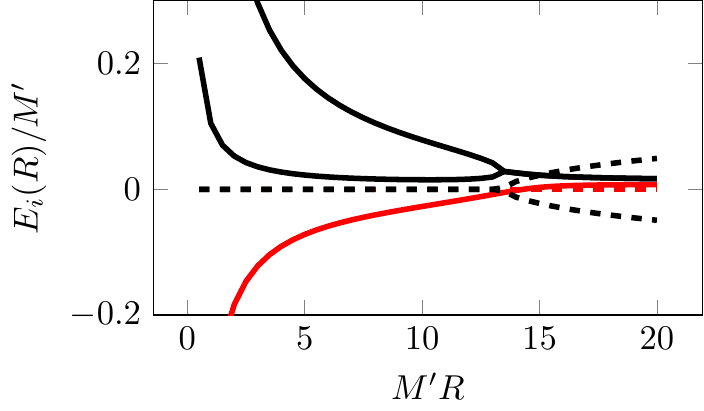}
  \caption{First three energy levels in the theory for $(\zeta',\zeta ) =(6.32,37.91)$. The solid lines represent the real part of the energy while the dashed lines represent the imaginary part. The ground state is highlighted in red.}
  \label{fig:002a02}
\end{subfigure}%
\hfill
\begin{subfigure}[t]{.45\textwidth}
  \centering
  \includegraphics[width=0.98\linewidth]{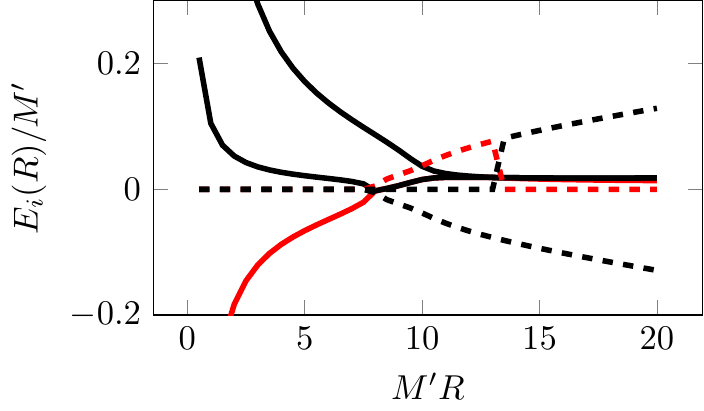}
  \caption{First three energy levels in the theory for $(\zeta',\zeta ) =(6.32,94.78)$. The solid lines represent the real part of the energy while the dashed lines represent the imaginary part. The ground state is highlighted in red.Here the ground state becomes real again for large volume: we are near the non-unitary tricritical point!}
  \label{fig:002a05}
\end{subfigure}
\caption{The typical behaviours observed in the tricritical Ising model observed for a larger positive value of $\zeta'$, at a level cut-off $N_\text{max}=10$.}
\end{figure}

For positive value of $\zeta'$ the same story repeats, but in reverse order: we pass from a region of values of $\zeta'$ in which for increasing values of $\zeta$ the ground state and the first exited state meet in a complex conjugate pair first (figure \ref{fig:00a001} and \ref{fig:00a05}) to higher values of $\zeta'$ such that for increasing values of $\zeta$ the first and the second excited states meet first  (figure \ref{fig:002a02} and \ref{fig:002a05}).
As in the previous case, this means that keeping $\zeta'>0$ and varying carefully the two parameter $\zeta$ and $\zeta'$ there should exist a value $(\zeta'_{c,2},\zeta_{c,2})$ of the couplings when the lowest three energy levels meet simultaneously. This is the point in the $v=0$ plane which is closest to the other the endpoints of the critical line that corresponds to the non-unitary tricritical Lee-Yang edge singularity. Again, to eventually hit the tricritical point requires switching on $v\neq 0$ and tuning it together with $\zeta$ and $\zeta'$ to put the triple meeting point towards (ideally) infinite volume.  

\begin{figure}[htb]
\centering
\begin{tikzpicture}[x=0.6pt,y=0.6pt,yscale=-1,xscale=1]
\draw[->] (200,10)--(200,-150);
\draw[->] (0,0)--(400,0);
\draw[dashed] (30,-150).. controls (170,-50).. (200,-40);
\draw[dashed] (350,-150).. controls (300,-50).. (200,-40);

\draw (210,-150) node [anchor=north west][inner sep=0.75pt]  [font=\large]  {$\zeta$};

\draw (400,0) node [anchor=north west][inner sep=0.75pt]  [font=\large]  {$\zeta'$};
\draw (110,-125) node [anchor=north west][inner sep=0.75pt]  [font=\large]  {A.};
\draw (172,-30) node [anchor=north west][inner sep=0.75pt]  [font=\large]  {B.};


\draw[blue] (30,-80) .. controls (70,-70) .. (120,-70);
\draw[blue] (30,-100) .. controls (70,-70) .. (120,-70);
\draw[blue] (30,-40) .. controls (70,-60) .. (120,-60);

\draw[blue] (185,-50) .. controls (240,-70) .. (280,-70);
\draw[blue] (180,-90) .. controls (240,-70) .. (280,-70);
\draw[blue] (180,-100) .. controls (230,-80) .. (280,-80);

\draw[blue] (330,-100) .. controls (370,-90) .. (420,-90);
\draw[blue] (330,-110) .. controls (370,-90) .. (420,-90);
\draw[blue] (330,-60) .. controls (370,-80) .. (420,-80);

\end{tikzpicture}
\caption{A cartoon in which the phenomenology observed from the energy spectrum in the $v=0$ plane is illustrated as a function of $(\zeta',\zeta)$. The blue sketches illustrate the observed phenomenology of the lowest three levels in the energy spectrum. Note that e.g. that when the ground state meets the first excited state in the blue drawings, we do not exclude that it also meets the second excited state for higher volumes, as in the case of Fig. \ref{fig:002a05}. The blue sketches only indicate the type of the meeting point that occurs first as $R$ is increased and correspond to phenomenology type A resp. B as discussed in the text. The dashed line separates the two different behaviours of the ground state and corresponds to the point in the plane in which the ground state meets the first excited state and the second excited state in the same point (phenomenology of type C). The lower half plane is related to the upper half by the symmetry with respect to the origin.}
\label{fig:PhenomenologyCartoon}
\end{figure}
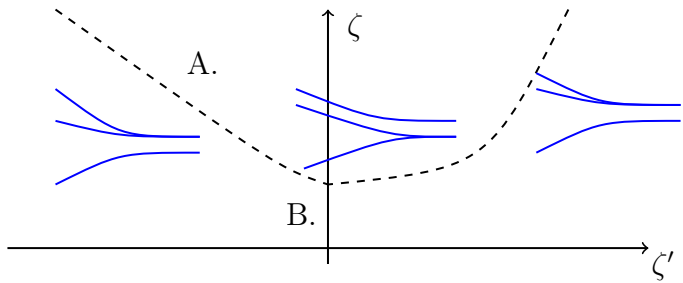

We note that the two points $(\zeta'_{c,1}, \zeta_{c,1})$ and $(\zeta'_{c,2}, \zeta_{c,2})$ are not related by any obvious symmetry, and that it also costs quite a high effort of fine-tuning to actually hit them (c.f. also the search for the eventual tricritical endpoint in the next subsection). The observed behaviour of the first three energy levels in the $v=0$ plane is summarised graphically in Fig. \ref{fig:PhenomenologyCartoon}. In particular the observed behaviours in the $v = 0$ plane are  
\begin{itemize}
    \item[A.] The ground state meets the first excited state, to form a complex conjugate pair, first. note that for a larges values of $R$ there could also be a meeting involving the second excited state.
    \item[B.] The first excited state meets the second excited state, to form a complex conjugate pair, first, then eventually the ground state meets the first excited state for higher values of $R$.
    \item[C.] The ground state meets for a certain finite value of $R$, the first and the second excited state simultaneously.
    \item[D.] The spectrum is entirely real.
\end{itemize}

The only point in the plane corresponding to real spectrum for every value of the radius $R$ (phenomenology of type D) is the origin, consistently with a spontaneously broken $\op P \op T$-phase in the whole plane (excluding the origin where unitarity is preserved)\footnote{Near the origin TCSA indicates that the first meeting point is between the first and the second excited states for a large value of $R$. However, this is impossible to demonstrate conclusively with TCSA since its range is limited the volume. Nevertheless, we do not have evidence of real spectrum in the plane and there are no reasons to think that a region in which the spectrum is completely real exists. If there exists a region dominated by the phenomenology of type D, it should be located in a small neighborhood of the origin, with its boundary composed by critical points of the Lee-Yang type. The detailed study of the existence of such a region is out of the scope of the present paper, since the focus of our interest is on the \textit{non-unitary tricritical points}.}. The TCSA results provide evidence for phenomenology of type $A$ (Figs. \ref{fig:zetap02}, \ref{fig:n0a03}, \ref{fig:n00a001}, \ref{fig:n00a02}, \ref{fig:00a001},\ref{fig:00a05}, \ref{fig:002a05}) and $B$ (Figs. \ref{fig:zeta01}, \ref{fig:zeta02}, \ref{fig:zetap01}, \ref{fig:n0a01}, \ref{fig:002a02}) separated by a line of points of phenomenology of type $C$.
 \newline

 \noindent The transition between the $\op P \op T$ unbroken/broken phases is expected to be governed by a critical point, as in the Ising case, however we have to recall that, in this section, we are discussing only the $v = 0$ plane and the full scaling region lives in three dimension (the three dimensional space spanned by $\zeta$, $\zeta'$ and the dimensionless parameter corresponding to $v$); then, from TCSA's results, it is clear that the $v = 0$ plane lives in the region of the theory space in which the $\op P \op T$ symmetry is always spontaneously broken (apart from the origin) and there are no critical points in it (since either the ground state meets always the first excited state for a finite value of $R$ or the first excited state meets the second excited state first).
This picture suggests that critical points of the type of \textit{non-hermitian critical point} (when the ground state meets the first excited state) and \textit{non-hermitian tricritical point} (when the ground state meets at the same time the first and the second excited state) are present in the scaling region when also $v$ is switched on. Therefore in the following we will look for this two classes of critical points in the full scaling region. 

\subsection{The non-unitary tricritical point}\label{TricriticalPoint}

In order to locate the non-unitary tricritical point it is necessary to consider the full scaling region, i.e. all couplings of the Hamiltonian \eqref{TricriticalHamiltonian} must be turned on. We note that the perturbation by the vacancy operator $t$ switched on by the coupling $v$ generates a divergent bulk energy contribution \cite{2011arXiv1106.2448G} and while this drops out from the relative energy levels, there is still an enhanced cut-off dependence as in the case of the Ising model. Fortunately it is still possible to attain the numerical precision which allows to determine the nature of the fixed point.

The fine tuning necessary to find the tricritical point is extremely difficult since it is necessary to make the lowest three levels meet at the same volume, while at the same time pushing the value of volume where they meet to as large as possible. Just as before, the independent dimensionless parameters are defined in terms of a scale constructed from the energy coupling $\mu$. For this subsection we choose the scale to be $M$ given by 
\begin{equation}
    \mu = \frac{10^{-1}}{2 \pi} M^{9/5}
\end{equation}
which locates the physical window for the tricritical fixed point conveniently in the interval $10\lesssim MR\lesssim 20$. 
\begin{equation}\label{DimensionlessParameters}
\xi = \frac{h}{M^{77/72}} \ , \hspace{1 cm} \xi' = \frac{h'}{M^{5/8}} \ , \hspace{1 cm} \tau = \frac{v}{M^{4/9}} \ .
\end{equation} 
The procedure of locating the tricritical point starts with finding a value for $(\xi,\xi')$ with $\tau=0$ where the three lines meet simultaneously at some specific value of the dimensionless volume parameter $MR$ and then switch on $\tau$ to try and push this volume as high as possible. During this process, however, the $(\xi,\xi')$ must be tuned together with $\tau$ to keep the three levels meeting at the same point. This results a very tedious process of searching for the optimal values of the couplings, and it is especially difficult to keep the three lines meet at the same volume. The search was performed by finding a rectangular box in $(\xi,\xi',\tau)$ space which definitely had the tricritical point inside, and subsequently halving sides of the box to get a subdivision into eight sub-regions and looking at the pattern of the spectrum in each of them, always selecting the box closest to the required phenomenology and then repeating the halving of the sides. 

\begin{figure}[b]
\begin{subfigure}[t]{.45\textwidth}
  \centering
  \includegraphics[width=0.98\textwidth]{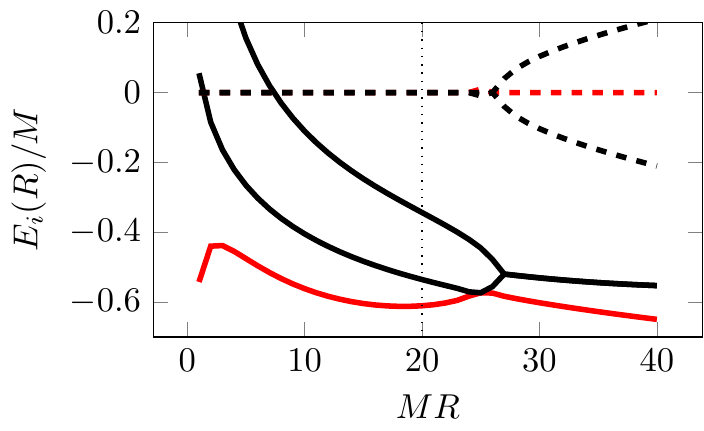}
  \caption{Solid/dashed lines depict the real/imaginary parts of the first three energy levels as a function of the volume at the couplings \eqref{NHTP}, with the ground state highlighted in red. The dotted vertical line indicates the end of the physical window.}
  \label{fig:tricRaw}
\end{subfigure}%
\hfill
\begin{subfigure}[t]{.45\textwidth}
  \centering
  \includegraphics[width=0.98\linewidth]{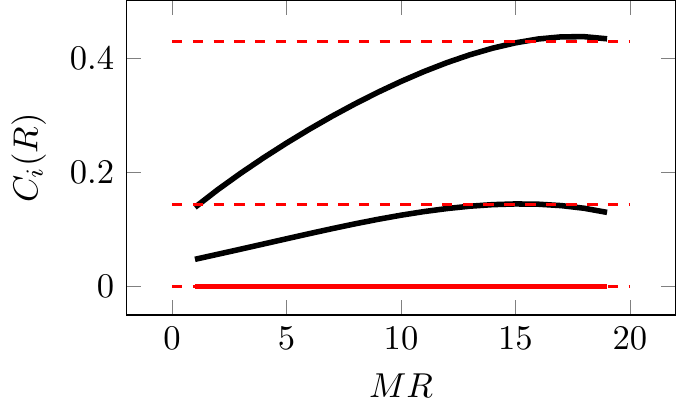}
  \caption{Solid lines show the scaling functions $C_i(R) = (E_i(R)-E_0(R))R/4 \pi$ corresponding to the real parts of the first three energy levels at the couplings in Eq. \eqref{NHTP}, while the red dashed lines are the predictions from the minimal model $\mathcal M(2,7)$.}
  \label{fig:tricC}
\end{subfigure}
\caption{The spectrum of the tricritical Ising model deformed with couplings tuned as in equation \eqref{NHTP} at level cut-off $N_\text{max}=10$. } 
\end{figure}

The search procedure described above ended up with the following location for the non-unitary tricritical point:
\begin{equation}\label{NHTP}
    \xi = 1.503(3) \,, \hspace{1 cm} \xi' = 0.119(4) \,, \hspace{1 cm} \tau = 0.512(1) \,.
\end{equation}
Fig. \ref{fig:tricRaw} illustrates the spectrum at these couplings. Note that it is eventually very hard to hit exactly the point where the three lines meet at the same value of the volume, but the subsequent analysis reveals that this point is eventually very close to the non-unitary tricritical point. 

In order to verify that the infrared CFT is the minimal model $\mathcal M(2,7)$ it is possible to compare the asymptotic values of the scaling functions $C_i(R) = (E_i(R)-E_0(R))R/4 \pi$ with the values expected from the minimal model $\mathcal M(2,7)$ which has three primary fields $\phi$, $\phi'$, and the identity $\op 1$ with conformal weights $(-3/7,-3/7)$, $(-2/7,-2/7)$  and $(0,0)$, respectively. The ground state corresponds to the primary field $\phi$, while for the first two excited state levels one obtains
\begin{equation}\label{eq: M27 prim}
\begin{split}
\phi': &\,C_1(R= \infty) =- \frac{2}{7}+ \frac{3}{7} =  \frac{1}{7}  \\
\op 1: &\,C_2(R= \infty)  = 0+ \frac{3}{7} = \frac{3}{7}
\end{split}
\end{equation} 
As shown in Fig. \ref{fig:tricC}, the match between these predictions and the TCSA results is quite convincing, at least for what concerns the lowest lines which correspond to the primary fields in the IR conformal field theory; the fourth and higher lines are descendant levels and, as discussed before, for these operators the errors due to truncation become relevant, in particular in our case, where the critical point is not really reached, but just approximated. A more detailed comment on the issue of higher energy levels can be found in appendix \ref{Appendix B}. This analysis tends then to confirm that the universality class of the infrared fixed point is indeed described by the minimal model $\mathcal M(2,7)$, in agreement with von Gehlen's results \cite{vonGehlen:1994rp}.

\subsection{Phenomenology on the critical line}

\begin{figure}[htb]
\begin{subfigure}[t]{.45\textwidth}
  \centering
  \includegraphics[width=0.98\textwidth]{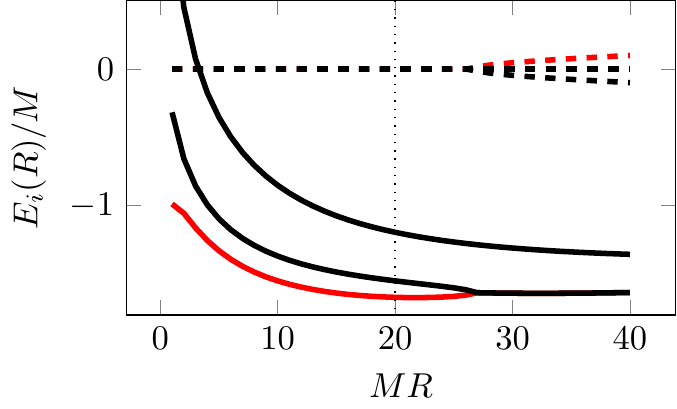}
  \caption{Solid/dashed lines depict the real/imaginary parts of the first three energy levels as a function of the volume at the couplings \eqref{NHCP}, with the ground state highlighted in red. The dotted vertical line indicates the end of the physical window.}
  \label{fig:CritRaw}
\end{subfigure}%
\hfill
\begin{subfigure}[t]{.45\textwidth}
  \centering
  \includegraphics[width=0.98\linewidth]{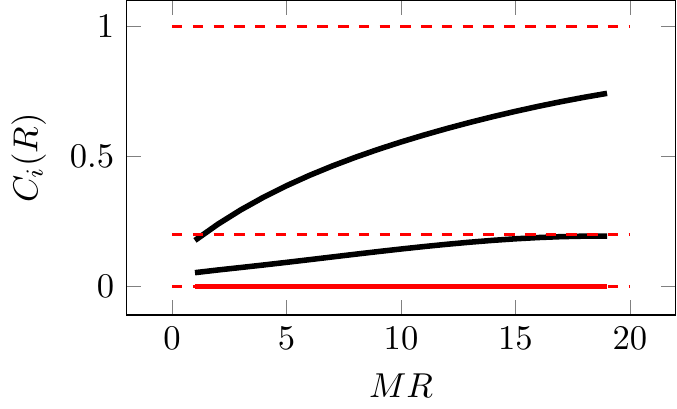}
  \caption{Solid lines show the scaling functions $C_i(R) = (E_i(R)-E_0(R))R/4 \pi$ corresponding to the real parts of the lowest lying energy levels at the couplings in Eq. \eqref{NHCP}, while the red dashed lines are the predictions from the minimal model.}
  \label{fig:critC}
\end{subfigure}
\caption{The spectrum of the tricritical Ising model deformed with couplings tuned as in equation \eqref{NHCP} at level cut-off $N_\text{max}=10$. } 
\end{figure}

While the most interesting of the fixed points reached by the RG flow induced by turning on imaginary magnetic fields in \eqref{TricriticalHamiltonian} is undoubtedly the tricritical Yang-Lee edge singularity, it is merely the endpoint of a line of critical points \cite{vonGehlen:1994rp}. Hitting the critical line is simpler than finding the tricritical end point since the number of parameters to be tuned is two instead of one. We find a point on the critical line by setting $h' = 0$ and tuning only $h$ and $v$  defined in equation \eqref{DimensionlessParameters}. The search in this case can be performed similarly as described for the critical point in the previous subsection, however this time it is much faster since it involves only two dimensions. The critical point is found to be located approximately at
\begin{equation}\label{NHCP}
    \xi = 55.93(3) \ , \hspace{1 cm} \xi' = 0 \hspace{1 cm} \tau = 7.62(0) \ , 
\end{equation}
with the corresponding energy spectrum shown in Fig. \ref{fig:CritRaw}. One can then evaluate the scaling functions $C_i(R) = (E_i(R)-E_0(R)) R/4 \pi$ and compare the results with the predictions following from the minimal model $\mathcal M(2,5)$ reported in Eq. \eqref{M25Predictions}. The dimension of the only nontrivial primary field is encoded in the scaling function $C_1(R)$ which should approach $1/5$ at the fixed point. As shown in Fig. \ref{fig:critC}, the numerical data agree with this prediction very well, confirming that the fixed points in the interior critical line indeed fall in the universality class of the minimal model $\mathcal M(2,5)$ i.e. the ordinary critical Yang-Lee singularity.

\section{Conclusions and outlook}\label{sec:Conclusions}

In this paper we have studied the tricritical extension of the Yang-Lee edge singularity by switching on imaginary magnetic fields in the tricritical Ising model in the framework of scaling quantum field theory. The infrared fixed points of the generated RG flow are interpreted in terms of the condensation of Yang-Lee zeros in the complex plane. 
The manifold of critical points is a line, where the internal points correspond to the ordinary critical Yang-Lee edge singularity with the corresponding universality class governed by the non-unitary minimal model $\mathcal M(2,5)$. We pointed out the crucial role played by $\op P\op T$-symmetry and its breaking, which is signalled by the merging of the lowest two levels into a complex conjugate pair in finite volume. The tricritical fixed points are the endpoints of the critical line and correspond to the three lowest levels meeting simultaneously in finite volume. Using the truncated conformal space approach it is possible to locate the critical line as well as the tricritical point, and extract the scaling weights of the primary fields in the infrared fixed point CFT. While at the critical line the numerical results are consistent with the infrared CFT $\mathcal M(2,5)$, the tricritical behaviour is controlled by the non-unitary minimal model $\mathcal M(2,7)$. These results are consistent with the $\op P\op T$-symmetric extension of Zamolodchikov's $c$-theorem \cite{Castro-Alvaredo:2017udm}, and are in full agreement with the results obtained on the lattice \cite{vonGehlen:1994rp}.

We expect that the same pattern holds also for higher multicritical points; in particular non-unitary tetracritical points are expected to result from switching on imaginary magnetic fields in $\mathcal M(5,6)$, while non-unitary pentacritical points should appear in $\mathcal M(6,7)$ and so on. We conjecture that the conformal minimal model that controls the $(n+1)$-th non-unitary multicritical point is the non-unitary minimal model $\mathcal M(2,2n+3)$, $n = 1,2, \ldots$. From a renormalisation group flows perspective this conjecture states that unitary minimal models $\mathcal M(p,p+1)$ perturbed with imaginary couplings for the odd fields produce, in the infrared, critical manifolds containing the minimal models $\mathcal M(2,2n+3)$, where $1 < n < p-1$. This picture is consistent with the $\op P\op T$-symmetric extension of Zamolodchikov's $c$-theorem, since $c_\text{eff}(\mathcal M(2,2n+3))<c_\text{eff}(\mathcal M(p,p+1))$ for every $n = 1, \ldots, 2p-1$ and $p\ge 3$. 

We remark that the number of nontrivial relevant fields in $\mathcal M(2,2n+3)$ is $n$, which is consistent with the tricritical points forming a boundary set of co-dimension $1$ of critical manifold, while the tetracritical points a boundary set of co-dimension $2$ and in the general the $n$-th multicritical points are expected to be a boundary set of the critical manifold with co-dimensions $n-2$. Note in the case of unitary multicritical points $\mathcal M(p,p+1)$ the number of nontrivial relevant fields  is $2(p-2)$ which naively seems twice the required number to specify the corresponding submanifolds. As explained for the tricritical Ising model in Subsection \ref{subsec:RG_imag_magnetic}, half of the relevant fields are eventually order parameters which are odd under $\mathbb Z_2$ (e.g. the leading and subleading magnetisations in the tricritical case), while the (multi)critical submanifolds are parameterised solely by the couplings of the even relevant fields. The counting of parameters is different for the multicritical Yang-Lee edge singularities since all fields in the Lagrangian must be included in the $\op P\op T$-symmetric deformations and there no analogues of order parameter fields which are present in the unitary case.

\acknowledgments

GM acknowledges the grant Prin 2017-FISI. The work of ML was supported by the National Research Development and Innovation Office of Hungary under the postdoctoral grant PD-19 No. 132118. GT and ML were partially supported by the Ministry of Culture and Innovation and the National Research, Development and Innovation Office (NKFIH) through the OTKA
Grant K 138606, and also within the Quantum Information National Laboratory of Hungary (Grant No. 2022-2.1.1-NL-2022-00004). ML, GM and GT are also grateful for the hospitality of KITP Santa Barbara during the  program "Integrability in String, Field, and Condensed Matter Theory", where a part of this work was completed. This collaboration was supported in part by the National Science Foundation under Grant No. NSF PHY-1748958, and by the CNR/MTA Italy-Hungary 2019-2021 Joint Project “Strongly interacting systems in confined geometries”. AM is grateful to SISSA, and to the Statistical Physics group in particular, for their warm hospitality during part of this work. 

\clearpage
\appendix

\section{Landau-Ginzburg description of the scaling region}\label{sec:AppendixA}

In this appendix we consider the extension of Zamolodchikov's argument relating unitary conformal minimal models and Landau-Ginzburg description of Ising multicritical points \cite{Zamolodchikov:1986db}, to the case of the non-unitary minimal models $\mathcal M(2,q)$.  

\subsection{Landau-Ginzburg description of unitary minimal models}
We start by recalling Zamolodchikov's original argument. The unitary conformal minimal models $\mathcal M(p,p+1)$ have conformal primary fields with dimensions: \begin{equation}\label{Primarydimunitary}
	\Delta_{r,s} = \frac{((p+1)r-ps)^2-1}{4 p(p+1)}\ , \hspace{1 cm} 1 \le r \le p-1 \  \ , 1 \le s \le p \ .
\end{equation}
The most relevant (nontrivial) field in the theory is the field $\phi_{2,2}$, which is expected to be identified with the order parameter $\varphi$ of the corresponding Landau-Ginzburg Lagrangian. The powers of the order parameter can then be computed by using the conformal operator product expansion. In particular
\begin{equation}
\begin{split}
	\varphi(x)\varphi(x') = \phi_{2,2}(x)\phi_{2,2}(x') = &  |x-x'|^{2 \Delta_{2,2}} \left(\op 1+\text{descendants}\right)+\\ & c_{\varphi \varphi}^{\phi_{3,3}} |x-x'|^{2 \Delta_{2,2}-\Delta_{3,3}} \left(\phi_{3,3}+\text{descendants}\right)+\\ & c_{\varphi \varphi}^{\phi_{1,3}} |x-x'|^{2 \Delta_{2,2}-\Delta_{1,3}} \left(\phi_{1,3}+\text{descendants}\right)+ \\ & c_{\varphi \varphi}^{\phi_{3,1}} |x-x'|^{2 \Delta_{2,2}-\Delta_{3,1}} \left(\phi_{3,1}+\text{descendants}\right) \ , 
	\end{split}
\end{equation} 
shows that after subtracting the identity contribution the renormalised square of $\varphi$ can be identified as 
\begin{equation}
: \varphi^2(0): = \lim_{x \to 0}|x|^{2 \Delta_{2,2}-\Delta_{3,3}}\left(\varphi(x)\varphi(0)-\langle \varphi(x)\varphi(0)\rangle\right) \sim \phi_{3,3}(0)\  .
\end{equation}
Iterating this process gives 
\begin{equation}\label{PowersOfOrderParameter}
: \varphi^k(0): = \lim_{x \to 0}|x|^{\Delta_1+ \Delta_{k-1}-\Delta_{k}} \left(\varphi(x):\varphi^{k-1}:(0)-\sum_{l= 1}^{k/2}|x|^{\Delta_{k-2l}-\Delta_1-\Delta_{k-1}}:\varphi^{k-2l}:(0)\right) \ .
\end{equation}
resulting in the identification 
\begin{equation}
    :\varphi^{k}: =
    \begin{cases}
    \phi_{k+1,k+1} & k=1,\ldots, p-2\\
    \phi_{k+3-p,k+2-p} & k=p-1,\ldots, 2p-4
    \end{cases}
\end{equation}
Note that the naive scaling dimensions of the Landau-Ginzburg field $\varphi$ and of all its powers are zero, and the positive scaling dimensions of the corresponding primary fields result from anomalous dimensions contributed by fluctuations. The scaling dimensions of the powers $\varphi^k$ are positive as required by unitarity, and they increase strictly monotonically with $k$.

All higher powers of $\varphi$ are identified with descendent fields. In particular,
considering the case $k=2p-3$ 
\begin{equation}
\begin{split}
\varphi(x) :\varphi^{2p-4}:(x') = &  c_{\varphi \varphi^{2p-4}}^{\varphi}|x-x'|^{\Delta_{\varphi^{2p-4}}}\left(\varphi(x)+\text{descendants}\right)+\\ & c_{\varphi \varphi^{2p-4}}^{\varphi^{2p-5}}|x-x'|^{\Delta_{\varphi}+\Delta_{\varphi^{2p-4}}-\Delta_{\varphi^{2p-5}}}\left(:\varphi^{2p-5}:(x)+\text{descendants}\right)\ ,
\end{split}
\end{equation}
where now both the $\varphi$ and $:\varphi^{2p-5}:$ must be subtracted, leaving us with
\begin{equation}
   \op L_{-1}\overline {\op L}_{-1} \varphi = c :\varphi^{2p-3}: \hspace{1 cm}\Leftrightarrow \hspace{1 cm}\partial \overline \partial  \varphi = c : \varphi^{2p-3}:\ ,
\end{equation}
for some numerical factor $c \in \mathbb R$. The above equation is the equation of motion of the Landau-Ginzburg theory with Lagrangian density is  
\begin{equation}
\mathcal L_\text{LG} = \frac{1}{2}\partial_\mu \varphi \partial^\mu \varphi +c' :\varphi^{2p-2}:\ .
\end{equation}
Furthermore, this correspondence implies that the scaling region of the Landau-Ginzburg theory is exactly spanned by the composite operators $:\varphi^{k}:$, $k=1,\dots,2p-4$ which are in one-to-one correspondence with the relevant primary fields in conformal field theory, completing the correspondence between the minimal model $\mathcal M(p,p+1)$ and the Landau-Ginzburg description of Ising multi-critical points.

\subsection{The case $\mathcal M(2,2n+3)$}
We now consider the extension of Zamolodchikov's argument to the case of non-unitary minimal models $\mathcal M(2,2n+3)$, $n = 1,2, \ldots$. These models contain $n+1$ primary fields with scaling dimensions 
\begin{equation}
\Delta_{1,r} = \frac{(2r-2n-3)^2-(2n+1)^2}{8 (2n+3)} \,  , \hspace{1 cm} 1 \le r \le n+1 \,.
\end{equation}
These are all negative which is related to the non-unitarity of the model. All the primary fields can be generated by taking successive operator powers of the field $\phi_{1,2}$ which is also the field whose scaling dimension is smallest in magnitude among all the nontrivial primary fields $\phi_{1,r}$ with $r\geq 2$. As a result it is natural to seek a Landau-Ginzburg description identifying $\phi_{1,2}$ as the fundamental field $\varphi$. Considering the operator product expansion  
\begin{equation}
\begin{split}
\varphi(x) \varphi(x') = \phi_{1,2}(x)\phi_{1,2}(x') = & |x-x'|^{2-6/(2n+3)}\left(\op 1+ \text{descendants}\right)+\\ & +c_{2,2}^{3} |x-x'|^{-2/(2n+3)}\left(\phi_{1,3}+ \text{descendants}\right)
\end{split}
\end{equation}
results in the identification $:\varphi^2: \sim \phi_{1,3}$, which by subsequent application of the operator product expansion $\phi_{1,2}\phi_{1,k}\sim\phi_{1,k-1}+\phi_{1,k+1}$ can be extended to the identification 
\begin{equation}
: \varphi^k: \sim   \phi_{1,k+1} \ , \hspace{1 cm} k = 2, \ldots,n \ .
\end{equation} 
In the last step one obtains
\begin{equation}\begin{split}
\phi_{1,2}(x)\phi_{1,n+1}(x')=& |x-x'|^{-1+1/(2n+3)} c_{1,n+1}^{n} \left(\phi_{1,n}(x)+\text{descendants}\right) \\  &  + |x-x'|^{-1+3/(2n+3)} c_{1,n+1}^{n+1} \left(\phi_{1,n+1}(x)+\text{descendants}\right) \ ,
\end{split}
\end{equation}
resulting in the equation of motion 
\begin{equation}\label{NonUnitaryGLEquation}
\op L_{-1} \overline{\op L}_{-1} :\varphi^{n}: = \gamma :\varphi^{n+1}: \,
\end{equation}
where $\gamma$ can be expressed in terms of the structure constants\footnote{Note that some of these structure constants are imaginary rather than real, again reflecting the non-unitary nature of the theory.} $c_{2r}^{r\pm 1}$. For the ordinary Yang-Lee edge singularity $n=1$, and the above reasoning reproduces the Landau-Ginzburg description described in Section \ref{sec:LY_zeros} which is captured by the Lagrangian \eqref{YLLagrangian} with an imaginary coupling reflecting the non-unitarity of the model.

For all $n\geq 2$, however, the equation of motion \eqref{NonUnitaryGLEquation} contains a non-canonical kinetic term. It can be cast into canonical form by expressing the dynamics in terms of the field $\rho = :\varphi^{n}:$; however in that the case the interaction is nonpolynomial corresponding to a Lagrangian
\begin{equation}
\mathcal L_{MF} = \frac{1}{2}\partial_\mu \rho \partial^\mu \rho +\gamma'  \rho^{2+1/n}\ .
\end{equation}
where $\gamma'$ is in general different from $\gamma$ which accounts for the multiplicative renormalisation of the field $\rho$. For $n\geq 2$, the physical interpretation of this  Lagrangian is not at all obvious and requires further investigation.
\section{First descendant level near the non-hermitian tricritical point}\label{Appendix B}

\begin{figure}[h!]
  \centering
  \includegraphics[width=0.7\textwidth]{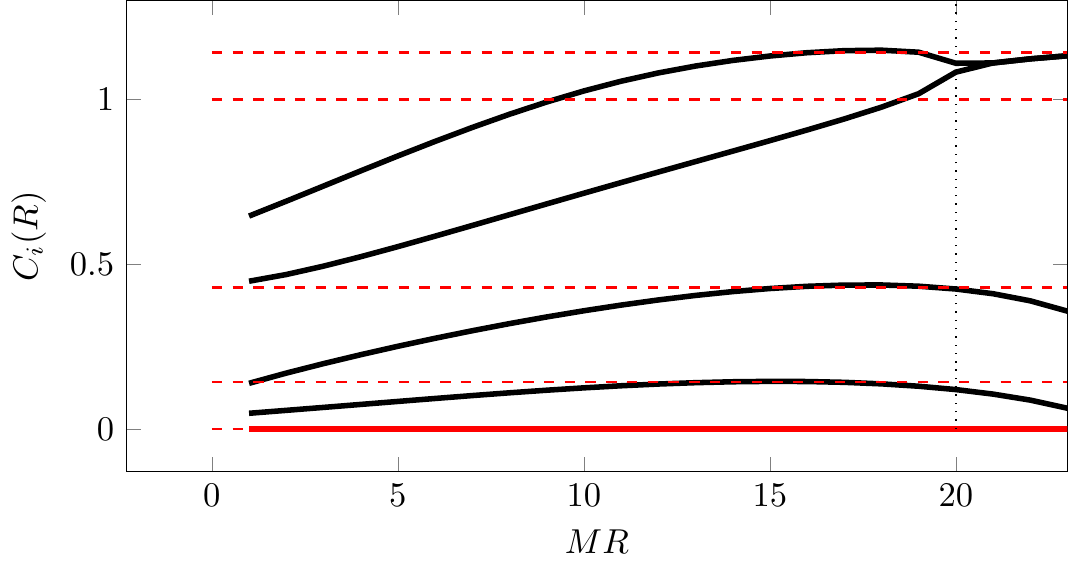}
  \caption{Solid lines show the scaling functions $C_i(R) = (E_i(R)-E_0(R))R/4 \pi$ corresponding to the real parts of the first five energy levels at the couplings in Eq. \eqref{NHTP}, while the red dashed lines are the predictions from the minimal model $\mathcal M(2,7)$.}
  \label{fig:tricD}
\end{figure}

In figure \ref{fig:tricC} we show that the differences between the first two excited states and the ground states, namely the quantities $C_1$ and $C_2$ defined in \eqref{eq: Ci definition}, for large volumes, reproduce the differences between the conformal dimensions of the minimal model $\mathcal M(2,7)$, as computed in equation \eqref{eq: M27 prim}. The TCSA allows us to compute higher energy levels and therefore it is interesting to see how these energy levels agree with their theoretical prediction for large volumes. In particular, hereafter we focus our  attention on those energy levels corresponding to the first descendent levels of the primaries $\phi$ and $\phi'$, that are expected to correspond to the third and fourth energy levels. For these levels, assuming the IR theory to be controlled by the minimal model $\mathcal M(2,7)$, one obtains
\begin{equation}
    \label{eq: predictions for Cdes}
    \begin{split}
\op L_{-1} \bar{\op L}_{-1}\phi: &\,C_3(R= \infty)  = \frac{4}{7}+ \frac{3}{7} = 1  \ ,\\
\op L_{-1} \bar{\op L}_{-1}\phi': &\,C_4(R= \infty)  = \frac{5}{7}+ \frac{3}{7} = \frac{8}{7} \ .
\end{split}
\end{equation}
The TCSA results are plotted in figure \ref{fig:tricD}, together with the predictions given in \eqref{eq: predictions for Cdes}.
Even if the agreement between the TCSA results and the predictions from the minimal model $\mathcal M(2,7)$ seems to be not accurate, the reader has to take in account that the quantities $C_i(R)$ converges to the expected values for $R\to \infty$ with different speeds. In the only known case, i.e. the Yang-Lee model $\mathcal M(2,5)$, obtained as a perturbation of the Ising $\mathcal M(3,4)$, the energies levels corresponding to the first descendant levels were plotted in \cite{Xu:2022mmw}. In that case the numerical method which was employed was the \textit{truncated free fermionic space approach} (TFFA), instead of the TCSA, and the convergence of the energy spectrum for large $R$ obtained by TFFA is expected to be much better. The reason is that, in the Ising case, the TFFA handles the energy perturbation exactly, leaving only the magnetic perturbation as numerical deformation. Nevertheless in that case the third and the fourth excited states meet each other at finite values of the radius $R$, splitting then, for larger values of $R$; such a splitting reduces the accuracy of the convergence of the third and fourth, which is still expected to occur for larger volumes. In the case of the non-hermitian tricritical point the situation seems to be similar: the third and the fourth excited states meet at finite volumes and seems to converge together at a value of $C$ which is very close to $C_4$. In the case considered in this paper unfortunately it is impossible to see the splitting of the two energy levels (as in the case of \cite{Xu:2022mmw}), since it is impossible to probe higher values of $R$, either because there are numerical errors due to TCSA (which become relevant for large value of $R$), and also for the fact that the non-hermitian tricritical point is never exactly not reached, but just approximated. Nothing to say but that future studies on the behaviour of $C_3$, $C_4$ and higher energy levels for larger volumes are desirable employing more powerful numerical methods than those used by us in this paper. 

\bibliographystyle{JHEP}
\bibliography{isingref}

\end{document}